\documentclass[traditabstract]{aa} 
\usepackage{graphicx}
\usepackage{latexsym}
\usepackage{color}
\usepackage{natbib}
\bibpunct{(}{)}{;}{a}{}{,} 
\usepackage{longtable,lscape}
\usepackage{multirow}

\newcommand{\teff}{$T_{\rm eff}$}
\newcommand{\logg}{$\log g$}
\newcommand{\vsini}{$v\sin i$}

\newcommand{\ebv}{$E(B-V)$}
\newcommand{\rotfit}{{\sf ROTFIT}}
\newcommand{\ROTFIT}{{\sf ROTFIT}}
\newcommand{\COMPO}{{\sf COMPO2}}
\newcommand{\kms}{km\,s$^{-1}$}
 
\newcommand{\gaia}{{\it Gaia}}

\definecolor{blu}{rgb}{0,0,1}
\definecolor{mag}{rgb}{1,0,1}

\begin{document}
\title{Stellar population astrophysics (SPA) with the TNG\thanks{Based on observations made with the Italian Telescopio Nazionale Galileo (TNG)
operated on the island of La Palma by the Fundaci\'on Galileo Galilei of the INAF (Istituto Nazionale di Astrofisica) at the Observatorio
del Roque de los Muchachos.
This study is part of the Large Program titled {\it SPA - Stellar Population Astrophysics:  the detailed, age-resolved
chemistry of the Milky Way disk} (PI: L. Origlia), granted observing time with HARPS-N and GIANO-B echelle spectrographs
at the TNG.}}
\subtitle{Characterization of the young open cluster \object{ASCC\,123}}

\author{A. Frasca\inst{1}\and 
	J. Alonso-Santiago\inst{1}\and
	G. Catanzaro\inst{1}\and
	A. Bragaglia\inst{2}\and
	E. Carretta\inst{2}\and
	G. Casali\inst{3}\and
	V. D'Orazi\inst{4}\and
	L. Magrini\inst{3}\and
	G. Andreuzzi\inst{5,6}\and
	E. Oliva\inst{3}\and
	L. Origlia\inst{2}\and
	R. Sordo\inst{4} \and
	A. Vallenari\inst{4}
	} 

\offprints{A. Frasca\\ \email{antonio.frasca@inaf.it}}

\institute{
INAF--Osservatorio Astrofisico di Catania, via S. Sofia 78, 95123 Catania, Italy
\and
INAF--Osservatorio di Astrofisica e Scienza dello Spazio, via P. Gobetti 93/3, 40129 Bologna, Italy
\and
INAF--Osservatorio Astrofisico di Arcetri,  Largo  E.  Fermi  5, 50125 Firenze, Italy
\and
INAF--Osservatorio Astronomico di Padova, vicolo dell'Osservatorio 5, 35122 Padova, Italy
\and
Fundaci\'on Galileo Galilei - INAF, Rambla Jos\'e Ana Fern\'andez P\'erez 7, 38712 Bre\~na Baja, Tenerife - Spain
\and
INAF--Osservatorio Astronomico di Roma, Via Frascati 33, 00078 Monte Porzio Catone, Italy}

\date{Received  / accepted}

 
\abstract{Star clusters are key to understand the stellar and Galactic evolution.
ASCC\,123 is a little-studied, nearby and very sparse open cluster. We performed the first high-resolution spectroscopic 
study of this cluster in the framework of the SPA (Stellar Population Astrophysics) project with GIARPS at the TNG.
We observed 17 stars, five of which turned out to be double-lined binaries. Three of the investigated sources were
rejected as members on the basis of astrometry and lithium content. 
For the remaining single  stars we derived the stellar parameters, extinction, radial and projected rotational velocities, 
and chemical abundances for 21 species with atomic number up to 40. From the analysis of single main-sequence stars we found an average 
extinction $A_V\simeq 0.13$\,mag and a median radial velocity of about $-5.6$\,\kms. The average metallicity we found for ASCC\,123 
is [Fe/H]$\simeq+0.14\pm 0.04$, which is in line with that expected for its Galactocentric distance.
The chemical composition is compatible with the Galactic trends in the solar neighborhood within the errors.
From the lithium abundance and chromospheric H$\alpha$ emission we found an age similar to that of the 
Pleiades, which agrees with that inferred from the Hertzsprung-Russell and color-magnitude diagrams.   } 

   \titlerunning{Stellar Population Astrophysics with the TNG. The young open cluster ASCC\,123}
      \authorrunning{A. Frasca et al.}

\maketitle

\section{Introduction}
\label{Sec:intro}

Star clusters play a fundamental role in astronomy. Stars formed in a cluster share a common origin, are located at the same distance (an 
approximation valid for all but the closest ones), and have an age and initial composition practically identical. However, small effects, as 
atomic diffusion, are observed in member stars of particularly well-studied clusters, as M67 \citep[e.g.,][]{Bertelli-Motta2018}. 
Therefore, they are the best laboratories to study 
stellar evolution and constrain theoretical models. Furthermore, the distribution of clusters of different age, mass, and distance allows to study
the Galactic disk. Open clusters (OCs) reveal the Galactic structure \citep[they have been used since the '60s to trace the spiral pattern of the 
Milky Way, see, e.g.,][]{Johnson1961,Janes1982,Junqueira2015} and permit to study the disk chemical evolution \citep[e.g.,][]{Friel1995}.

SPA (Stellar Population Astrophysics) is an on-going Large Program running on the $Telescopio$ $Nazionale$ $Galileo$ (TNG) at La Palma. 
SPA is an ambitious project which aims to perform an age-resolved chemical map of the Solar neighborhood and the 
Galactic thin disc. 
More than 500 representative stars (most of them located in clusters), covering different distances, ages and evolutionary stages, will be observed in the 
optical and near-infrared bands (NIR) at high resolution by combining HARPS-N ($R=115\,000$) and GIANO-B ($R=50\,000$) spectrographs 
\citep[see][for more details on SPA]{Origlia2019}.
The detailed chemical tagging carried out in this project will be combined with astrometric and photometric data obtained by the $Gaia$ mission. This 
will provide many valuable clues to improve our understanding of possible trends, gradients, age-metallicity relations, which allow us to 
constrain stellar evolution models as well as to determine in an accurate way the chemical evolution of the Galaxy. 
This work is one of the first papers of a series devoted to present the results of the SPA project. 

ASCC\,123 is a little-studied cluster located in the second Galactic quadrant [$\alpha$(2000)\,=\,$22^{\rm h}42^{\rm m}35^{\rm s}$, 
$\delta$(2000)\,=\,+54$^{\circ}$15$'$35$''$, $\ell$\,=\,104.74$^{\circ}$, $b$\,=\,$-4.00^{\circ}$].
It was discovered by \citet{Kharchenko2005} when analyzing the $Hipparcos$ proper motions and $BV$ archival photometry contained in the All-Sky Compiled 
Catalog of 2.5 million stars \citep[ASCC-2.5,][]{Kharchenko2001}. 
They revealed the existence of a new cluster composed of 24 likely members spread in a large region on the sky with a radius of about 77$\arcmin$. 
They found a small reddening, $E(B-V)=0.10$, and placed the cluster at a distance of 250 pc. They also estimated, with a large uncertainty, an age of 260 Myr and a radial 
velocity, $RV$\,=\,$-6.5$\,km\,s$^{-1}$.
Later, by combining astrometric data and near infrared photometry from the PPMXL and 2MASS catalogs, \citet{Kharchenko2013} refined the reddening, $E(B-V)=0.15$, 
and the age ($\tau=155$~Myr) of the cluster. 
\citet{Yen2018}, based on the $Gaia$ DR1/TGAS and HSOY \citep{HSOY}, reanalyzed the parameters of the cluster finding values compatible with the previous ones: 
$E(B-V)=0.097$, $d$=243.5 pc and an age of 130 Myr.
Finally, \citet{Cantat2018} from the $Gaia$ DR2 astrometric data estimated a likely distance of $233.1\pm5.5$\,pc. These works computed the cluster extent from different 
approaches, which makes their comparison difficult. All estimates show a large and diffuse cluster with a core radius between 6--17$\arcmin$ which extends up to, 
at least, 60--78$\arcmin$. Depending on the radius considered, the cluster hosts from 24 to 121 stars as most likely members. 

This is the first paper devoted exclusively to ASCC\,123 so far. Until now, it has been studied in an automatic way along with many more clusters. 
In this work we provide high-resolution spectroscopy for 17 candidate members of this cluster for the first time with the aim
of performing a full characterization in terms of atmospheric parameters, elemental abundance, rotation velocity, 
and activity level.

The paper is organized as follows. We present our data in Sect.~\ref{Sec:Observations}, the cluster membership in Sect.~\ref{Sec:membership}, 
and color-magnitude diagrams in Sect.~\ref{Sec:CMD}.
The data analysis and the main results are presented in Sect.~\ref{Sec:anal} regarding the stellar parameters, and in Sect.~\ref{Sec:chrom_lithium} regarding the
lithium content and chromospheric activity. The global properties of ASCC\,123 are discussed in Sect.~\ref{Sec:disc}. 
Section~\ref{Sec:summary} summarizes the results of this work.

%
\section{Observations and data reduction}
\label{Sec:Observations}

In order to study the cluster we selected a series of representative stars among the brightest likely members according to the literature. 
The selection was done before the publication of the second \textit{Gaia} data release. Therefore, as a first step we started 
with the list of 959 stars proposed by \citet{Kharchenko2005,Kharchenko2013} as potential candidates. 
The selection was refined, rejecting several Kharchenko's candidates and adding two stars, with the new \textit{Gaia} DR2 data.	

The observations were conducted from 18 to 22 August 2018 and on 11 August 2019 (see Table~\ref{Tab:obs_log}) with GIARPS 
(GIANO-B \& HARPS-N, \citealt{Claudi2017}) at the 3.6-m TNG telescope located at El Roque de los Muchachos Observatory, in La Palma (Spain).
GIARPS uses both the optical high-resolution spectrograph
HARPS-N ($R\simeq 115\,000$, range\,=\,0.39--0.68\,$\mu$m, \citealt{Cosentino2014}) and the near-infrared spectrograph GIANO-B 
($R\simeq 50\,000$, range\,=\,0.97--2.45\,$\mu$m, \citealt{oli12a,oli12b,ori14}).
The HARPS-N spectra were acquired with the second fiber on-sky.
The spectra were acquired with total exposure times ranging from 1200 to 5700 sec, depending on the star brightness and sky conditions, with the aim 
of reaching a signal-to-noise ratio per pixel SNR$\geq$\,30 at red wavelengths. Exposures longer than 1800 sec were usually split 
in two or three to reduce the contamination of cosmic rays.
The properties of the 17 stars observed at the TNG are listed in Table~\ref{Tab:Targets}.
In the present paper we make only use of HARPS-N spectra, which are best suited for the determination of stellar parameters and abundances of 
these relatively faint sources.

The HARPS-N spectra were reduced by the instrument Data Reduction Software pipeline. Radial velocities (RVs) were derived by this pipeline 
using the weighted cross-correlation function (CCF) method \citep{Baranne1996, Pepe2002}. 
However, we have redone the CCF analysis using synthetic template spectra and a broader RV window, because most of our targets are binaries or 
rapidly rotating stars whose CCF peaks were not entirely covered by the RV range of the online CCF procedure. The RV values are reported in 
Table~\ref{Tab:Spectra_param}. 

The telluric H$_2$O lines at the H$\alpha$ and Na\,{\sc i}\,D$_2$ wavelengths, as well as those of O$_2$ at 6300 \AA, were removed from the
extracted HARPS-N spectra using an interactive procedure described by \citet{Frasca2000} and adopting telluric templates (spectra of hot, 
fast-rotating stars) acquired during the observing run. 

\setlength{\tabcolsep}{4pt}

\begin{table}
\caption{Observation log.}
\begin{center}
\begin{tabular}{lccccr}   
\hline\hline
\noalign{\smallskip}
\multirow{2}{*}{Id$^{a}$} & \multirow{2}{*}{Name}   &  Date\_obs                & UT\_mid             & $t_{\rm exp}$ & \multirow{2}{*}{SNR}$^b$ \\	       
                          &                         &  {\scriptsize yyyy-mm-dd} & {\scriptsize hh:mm} &    (s) 	      &	                     \\
\hline
\noalign{\smallskip}
39  & \object{BD+54\,2812}     &   2018-08-19  &  02:19 & 3600  &  106  \\
56  & \object{HD 235888}       &   2018-08-19  &  05:15 & 1183  &  125  \\
56  & \object{HD 235888}       &   2018-08-21  &  04:15 & 1410  &  147  \\
214 & \object{TYC 3983-2832-1} &   2018-08-20  &  02:36 & 5700  &   90  \\
266 & \object{BD+52\,3260}     &   2018-08-19  &  03:28 & 3600  &  111  \\
378 & \object{TYC 3984-1809-1} &   2018-08-24  &  00:21 & 2820  &  153  \\
435 & \object{TYC 3988-1537-1} &   2018-08-23  &  05:04 & 5700  &   81  \\
466 & \object{HD 215178}       &   2018-08-20  &  01:15 & 3600  &  419  \\
490 & \object{TYC 3984-1751-1} &   2018-08-23  &  00:03 & 3600  &   70  \\
492 & \object{TYC 3988-154-1}  &   2018-08-23  &  23:19 & 3600  &  109  \\
502 & \object{TYC 3988-282-1}  &   2018-08-20  &  22:22 & 3600  &   75  \\  
517 & \object{TYC 3984-1588-1} &   2018-08-20  &  04:44 & 5700  &   80  \\
554 & \object{TYC 3984-1107-1} &   2018-08-23  &  01:08 & 3600  &  132  \\
565 & \object{TYC 3984-1717-1} &   2018-08-22  &  01:47 & 3600  &  121  \\
708 & \object{TYC 3984-277-1}  &   2018-08-19  &  04:31 & 3600  &  153  \\
731 & \object{TYC 3988-1310-1} &   2018-08-22  &  02:41 & 1410  &  105  \\
731 & \object{TYC 3988-1310-1} &   2018-08-23  &  01:52 & 1380  &  110  \\
F1  & \object{J22452826+5347061} & 2019-08-11  &  00:50 & 4800  &   36  \\
F2  & \object{J22311798+5502407} & 2019-08-11  &  02:23 & 4800  &   34  \\
\noalign{\smallskip}
\hline  
\end{tabular}
\begin{list}{}{}
\item[$^a$] Sequential number from \citet{Kharchenko2005} based on right ascension. 
F1 and F2 are two further candidates not included in \citet{Kharchenko2005}.
\item[$^b$] Signal-to-noise ratio per pixel at 6500\,\AA.
\end{list}
\label{Tab:obs_log}
\end{center}
\end{table}

\subsection{Archival data}
In order to complement our spectroscopic observations, as well as to characterize other likely cluster members, we resorted to archival data provided
by some all-sky surveys. Specifically, we took $BV$ optical photometry from the APASS \citep{APASS} and TYCHO \citep{HIPPA97} catalogs for the fainter 
and brighter sources, respectively. $I_{\rm C}$ magnitudes were retrieved from the TASS catalog \citep{TASS} and near infrared $JHK_{\textrm{s}}$ magnitudes from the 
2MASS catalog \citep{2MASS}. 

In addition, we also took advantage of the great possibilities offered by the second $Gaia$ data release \citep{GaiaDR2}. 
We collected the high-quality photometric and astrometric data available in this release for the stars under study.

The photometric data are listed in Table~\ref{Phot_123}, while the astrometric properties are reported in Table~\ref{Tab:Targets}.

\begin{table*}
\caption{Astrometric data of the stars in the field of ASCC\,123 observed at the TNG.} 
\begin{center}
\begin{tabular}{lcccrcccccrc}   
\hline\hline
\noalign{\smallskip}
\multirow{2}{*}{Id$^{a}$}  & \multirow{2}{*}{\gaia\ DR2 Id} &   RA    &  DEC    &  $r$~~	    & \multirow{2}{*}{$P_{\textrm{Kh05}}^{b}$} & \multirow{2}{*}{$P_{\textrm{Ca18}}^{c}$} & $\varpi$ &  $\mu_{\alpha*}$ &  $\mu_{\delta}$ & $G$~~ & $B_{\rm P}$--$R_{\rm P}$\\    
                           &  & \scriptsize{(J2000)} & \scriptsize{(J2000)} & \scriptsize{(\arcmin)}~~  &	    &	  & \scriptsize{(mas)} & \scriptsize{(mas\,yr$^{-1}$)} & \scriptsize{(mas\,yr$^{-1}$)} & \scriptsize{(mag)} & \scriptsize{(mag)} \\       
\noalign{\smallskip}
\hline
\noalign{\smallskip}
\multicolumn{10}{c}{Members}\\
\noalign{\smallskip}
\hline
\noalign{\smallskip}
39   & \scriptsize{2003378188041736320} & 22 35 13.26  & +54 46 24.8 & 71.1 &  0.80  &  1.00 &  4.1879 & 12.303 &  $-$0.578  & 10.352 &  0.624  \\    
56   & \scriptsize{2003324655567305728} & 22 35 36.32  & +54 32 17.5 & 63.2 &  0.93  & \dots &  4.0747 & 13.236 &  $-$0.054  &  9.052 &  0.574  \\  	  
214  & \scriptsize{2003023629898711680} & 22 38 34.03  & +53 35 08.7 & 53.8 &  0.69  &  1.00 &  4.2053 & 12.187 &  $-$1.355  & 11.830 &  0.925  \\  
266  & \scriptsize{2002270842392282496} & 22 39 19.22  & +53 29 16.5 & 54.6 &  0.83  &  1.00 &  4.1439 & 11.842 &  $-$1.381  & 10.599 &  0.766  \\     
378  & \scriptsize{2003132344109912832} & 22 41 05.97  & +53 59 04.2 & 21.0 &  0.40  &  0.70 &  4.2965 & 11.309 &  $-$1.324  &  9.548 &  0.421  \\  
435  & \scriptsize{2003443437181978240} & 22 42 00.19  & +55 00 58.5 & 45.7 &  0.99  &  1.00 &  4.2698 & 12.201 &  $-$1.388  & 12.050 &  0.952  \\  
466  & \scriptsize{2003171651651105664} & 22 42 24.04  & +54 14 54.2 &  1.7 &  0.61  &\dots  &  4.2145 & 13.523 &  $-$2.111  &  7.550 &  0.081  \\    
492  & \scriptsize{2003173713235666560} & 22 42 51.15  & +54 23 53.5 &  8.6 &  0.81  &\dots  &  4.7671 & 10.785 &  ~~0.102   & 10.635 &  0.883  \\   
517  & \scriptsize{2003161751738142464} & 22 43 26.53  & +54 11 58.4 &  8.4 &  0.70  &  1.00 &  4.2932 & 12.198 &  $-$1.831  & 12.012 &  1.010  \\  
554  & \scriptsize{2002409483936262016} & 22 44 00.20  & +54 08 38.1 & 14.3 &  0.95  &  1.00 &  4.1943 & 12.285 &  $-$1.693  & 10.171 &  0.643  \\  
708  & \scriptsize{2002765416464629504} & 22 46 15.85  & +54 01 36.6 & 35.2 &  0.75  &\dots  &  4.6208 & 11.393 &  $-$1.470  & 10.015 &  0.753  \\   
731  & \scriptsize{2003554006818945280} & 22 46 34.54  & +54 46 05.2 & 46.2 &  0.97  &  1.00 &  4.3944 & 12.589 &  $-$1.827  &  9.573 &  0.451  \\ 
F1   & \scriptsize{2002337603362057088} & 22 45 28.25  & +53 47 06.1 & 38.2 & \dots  &  1.00 &  4.2593 & 11.937 &  $-$1.489  & 12.567 &  1.065  \\ 
F2   & \scriptsize{2006435105245732480} & 22 31 17.98  & +55 02 40.7 &108.6 & \dots  &  1.00 &  4.1186 & 11.758 &  $-$0.888  & 12.765 &  1.093  \\ 
\noalign{\smallskip}
\hline
\noalign{\smallskip}
\multicolumn{10}{c}{Non members}\\
\noalign{\smallskip}
\hline 
\noalign{\smallskip}
490 & \scriptsize{2002403436622121472} & 22 42 49.48  & +54 03 11.9 & 12.6 &  0.21  & \dots   &  3.6143 &  9.488 &  $-$5.986  & 11.377 &  0.780  \\   
502 & \scriptsize{2003178592318525696} & 22 43 03.00  & +54 22 53.8 &  8.4 &  0.79  & \dots   &  4.5700 & 10.085 &  $-$2.686  & 11.020 &  0.961  \\   
565 & \scriptsize{2002120243658959488} & 22 44 05.71  & +53 13 56.3 & 63.1 &  0.79  & \dots   &  4.5545 & 12.191 &   ~~2.839  & 10.339 &  0.676  \\ 
\noalign{\smallskip}
\hline  
\end{tabular}
\begin{list}{}{}
\item[$^a$] Star identifier as defined in Table~\ref{Tab:obs_log}. 
\item[$^b$] Membership probability according to \citet{Kharchenko2005}, based on Hipparcos astrometry.
\item[$^c$] Membership probability according to \citet{Cantat2018}, based on \gaia\ DR2 astrometry.
\end{list}
\label{Tab:Targets}
\end{center}
\end{table*}

\section{Cluster membership}
\label{Sec:membership}

\begin{figure} 
\hspace{-.7cm}
  \includegraphics[width=9.8cm]{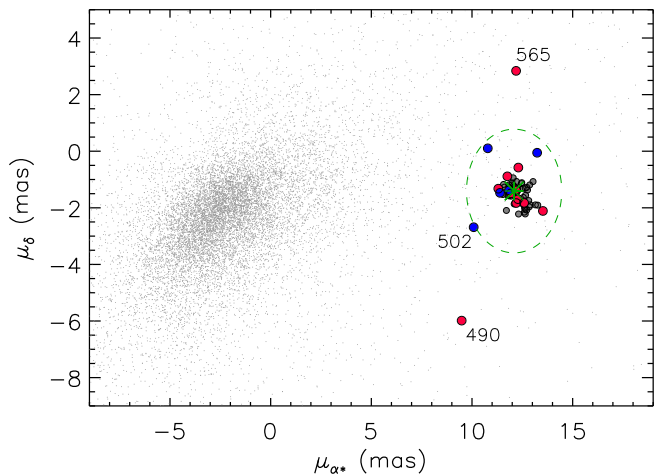}  
  \caption{Proper motion diagram of the sources with $G\leq 15$\,mag in the ASCC\,123 field (center coordinates 
  $\alpha$(2000)\,=\,$22^{\rm h}42^{\rm m}35^{\rm s}$, $\delta$(2000)\,=\,+54$^{\circ}$15$'$35$''$, radius $75\arcmin$). 
The stars investigated in the present paper are marked with big dots (red for single stars, blue for SB2 systems).
The candidate members according to \citet{Cantat2018} are represented with gray filled circles; the green asterisk denotes the average proper 
motion of ASCC\,123 according to \citet{Cantat2018} and the dashed ellipse is the 5-$\sigma$ contour. The three non-members, which lie outside 
this locus are also labeled with their Id.}
  \label{fig:ppm} 
\end{figure}

The disentangling of cluster members from field stars is not a trivial task and it is key to characterize the cluster itself.
Based on the analysis of the proper motions from $Hipparcos$, \citet{Kharchenko2005} identified 48 likely members to which they assigned a membership 
probability ($P$) greater or equal to 0.6. These stars are distributed in a wide region of the sky centered at about $\alpha$(2000)\,=\,$22^{\rm h}42\fm5$, 
$\delta$(2000)\,=\,$+54\degr 15\arcmin$ with a radius of $\sim 75 \arcmin$.
\citet{Cantat2018}, taking advantage of the very precise astrometry provided by $Gaia$ DR2, identified 55 members, considered as those stars with 
$P\geq 0.6$ in a field covering up to 140$\arcmin$ from the cluster center. For our targets the membership probabilities from the two aforementioned works 
are listed in Table\,\ref{Tab:Targets}.
Seven stars in our sample, namely S\,39, S\,214, S\,266, S\,435, S\,517, S\,554  and S\,731, are likely members according to both works. On the contrary, 
star 378 is quoted as a likely member in \citet{Cantat2018} only, whereas six other objects have a membership probability $P_{\textrm{Kh05}}\geq 0.6 $ 
according to \citet{Kharchenko2005}, but have no entry in the Cantat Gaudin catalog. Three of them  (S\,56, S\,492, and S\,708) turned out to be 
double-lined spectroscopic binaries (SB2s). Their position is close to the 5-$\sigma$ contour in the proper motion diagram of Fig.~\ref{fig:ppm}.
The stars labeled as F1 and F2 are not included in the Kharchenko's list, but they are very likely members ($P_{\rm Ca18}$\,=\,1) according to \citet{Cantat2018}. 
Only S\,490 is not a member in both works. In fact, its astrometric parameters are clearly different from 
the average values of the cluster as evaluated with the remaining stars.
Moreover, as we will show later (see Sect.\,\ref{Sec:chrom_lithium}), its very low lithium abundance is not compatible with the cluster age. In addition, 
S\,502, despite the high value of membership probability ($P_{\textrm{Kh05}}=0.79$), can not be accepted as a cluster member because of its very 
low lithium abundance. 

\begin{figure} 
\hspace{-.7cm}
  \includegraphics[width=9.8cm]{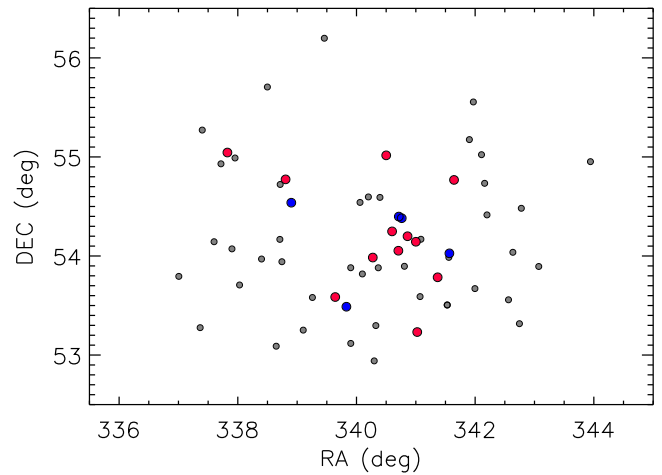}  
  \caption{Spatial distribution of the sources investigated in the present paper (big red dots for single stars, big blue dots for SB2 systems).
The candidate members according to \citet{Cantat2018} are represented with gray filled circles. }
  \label{fig:spatial} 
\end{figure} 

The spatial distribution of the stars investigated in the present paper is shown in Fig.~\ref{fig:spatial}, where the members 
according to \citet{Cantat2018} are also shown with smaller gray dots.
The astrometric properties of our targets are displayed in the $\mu_{\alpha*}$--$\mu_{\delta}$ diagram (Fig.~\ref{fig:ppm}) along with the
members according to \citet{Cantat2018} and the stars in the field of ASCC\,123 with magnitude $G\le 15$\,mag.
\citet{Cantat2018} computed mean proper motions for ASCC\,123 of $\overline{\mu_{\alpha*}}=12.093$ and $\overline{\mu_{\delta}}=-1.407$\,mas\,yr$^{-1}$, 
with standard deviations $\sigma_{\mu_{\alpha*}}=0.473$ and $\sigma_{\mu_{\delta}}=0.437$\,mas\,yr$^{-1}$.  We adopted a 5-$\sigma$ criterion, shown by
the ellipse in  Fig.~\ref{fig:ppm}, to confirm or discard candidate members. 
We found that, in addition to S\,490, also S\,565 is located well outside this boundary and should be considered as a non-member. It has the 
largest RV among the single-lined objects ($RV=-1.8$\,\kms), which supports its non-membership. However, as we shall see later, its \teff$>6500$\,K does not permit 
to use the lithium abundance as a further criterion for membership. The other non-member, S\,502, is just out of the 5-$\sigma$ boundary in
the proper motion diagram.

In summary, among the 17 stars forming our sample, 14 are likely cluster members, whereas the other three, S\,490, S\,502, and S\,565, 
should be considered as non-members.

To reinforce or reject the membership of our targets to ASCC\,123, additional diagnostics based on high-resolution spectra, such as radial 
velocity, chromospheric activity, and lithium atmospheric content, must be used. This analysis is presented in Sects.~\ref{Sec:anal} and 
\ref{Sec:chrom_lithium}.

\section{Color-magnitude diagrams and isochrone fitting}
\label{Sec:CMD}

\begin{figure*}  
  \centering         
  \includegraphics[width=19cm]{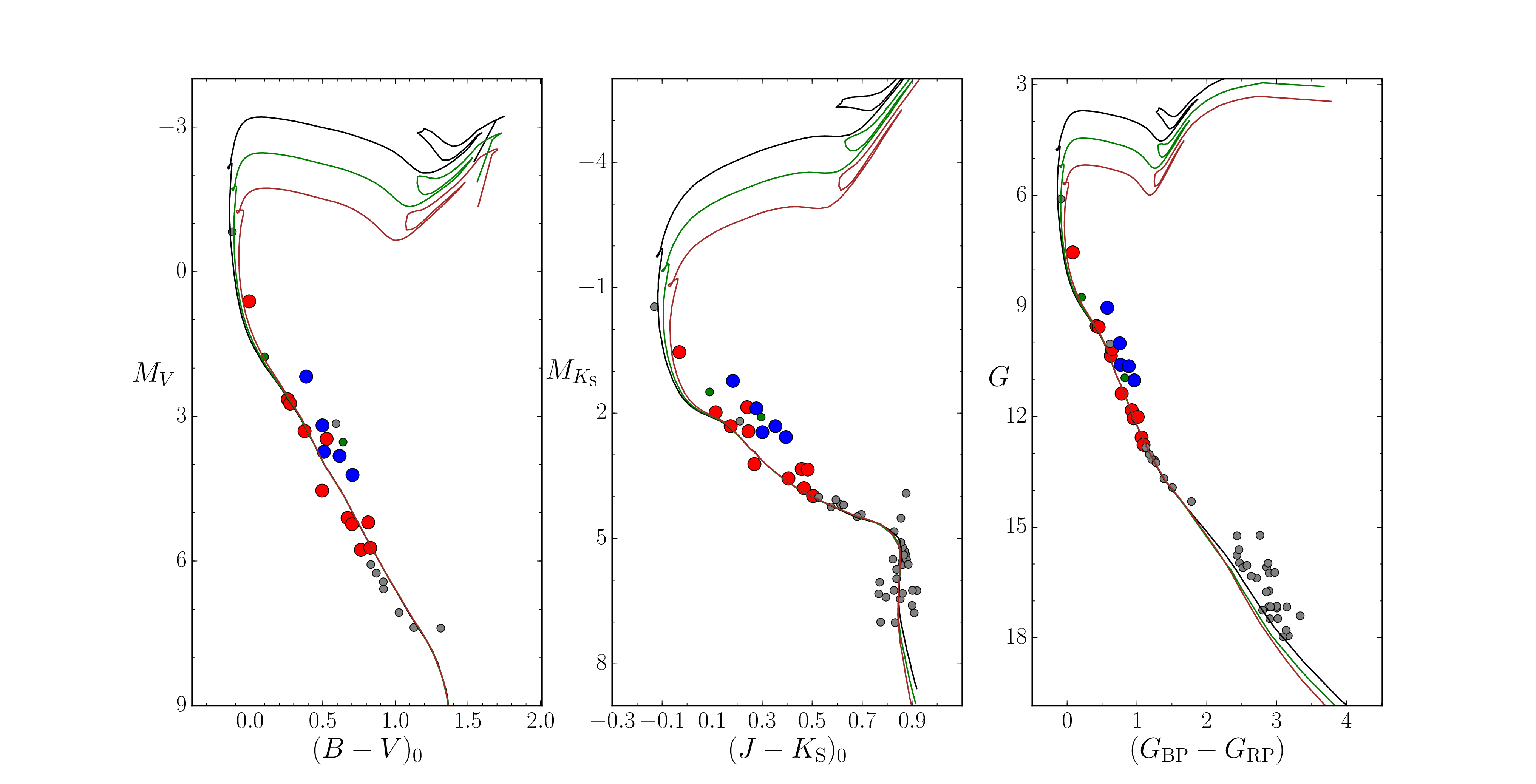}
  \caption{Color-magnitude diagrams for likely members of ASCC\,123 in three different photometric systems: \textit{Left:} $M_V/(B-V)_0$ from the APASS catalog;  
  \textit{Center:} $M_{K_{\textrm{s}}}/(J-K_{\textrm{s}})_0$ (2MASS photometry) and \textit{Right:} $G/(G_{\rm BP}-G_{\rm RP})$ ($Gaia$ DR2 data). Photometric data for likely 
  members appear as gray dots.
  Stars observed spectroscopically are represented as red circles (single or possible SB1 objects) or blue circles (SB2). 
  The solid lines represent the best-fitting PARSEC isochrones computed at  $\tau=100$\,Myr (black line), $\tau=155$\,Myr (green line), and 
  $\tau=250$\,Myr (red line) assuming the  metallicity and extinction found in the present work and the distance $d=233$\,pc \citep{Cantat2018}. }
  \label{fig:isoc}  
\end{figure*}

The most common procedure to estimate the age of a star cluster is the so-called isochrone-fitting method. On a color-magnitude diagram (CMD) several 
isochrones computed at different ages are drawn to find the one that best reproduces the location of the members on the diagram.
While automated or Bayesian-based methods are well suited for most clusters, ASCC\,123 was not included either in \citet{Bossini2019} or in
\citet{Monteiro2019}. So, in order to ensure the reliability of the determination of the cluster parameters, we selected only those stars whose membership 
probability according to \citet{Cantat2018} is sufficiently high 
(i.e. $P_{\rm Ca18}\geq$\,0.75) as well as those members from \citet{Kharchenko2005} whose \gaia\ DR2 astrometry is compatible with the cluster.
In Fig.\,\ref{fig:isoc} three CMDs are plotted in different photometric systems: optical $M_V/(B-V)_0$ (left panel), infrared 
$M_{K_{\textrm{s}}}/(J-K_{\textrm{s}})_0$ (central panel) and $Gaia$ magnitudes $G/(G_{\rm BP}-G_{\rm RP})$ (right panel). On these CMDs we overplotted the 
targets observed in this work, distinguishing single from SB2 stars.
Then we added PARSEC isochrones \citep{Bressan2012} computed at the average metallicity found in 
this work, [Fe/H]=+0.14 (see Sect.\,\ref{subsec:abundance}).
In the cases of the optical and infrared photometry we corrected the magnitudes and colors for interstellar dust extinction adopting 
the average value found in the present paper for the single stars, $A_V=0.13$ (Sect.\,\ref{subsec:HR}).		
In the \gaia\ CMD we have instead preferred to display the reddened isochrones, computed with the calibrations of \citet{MaizApellaniz18}, since the 
dereddening of the $Gaia$ photometry is not a trivial task. 
The distance adopted was that reported by \citet{Cantat2018} from $Gaia$ parallaxes, i.e. $d=233$\,pc.

In all three CMDs displayed in Fig.~\ref{fig:isoc}, three isochrones computed at 100, 155, and 250 Myr are plotted. The fit is quite good and, 
as expected, for the same color the SB2s are brighter than single stars. The absence of turn-off and evolved stars prevents an accurate determination of the 
cluster age. This is because in the mid and upper main sequence (MS), where our members are located, isochrones in a wide age range behave in the same manner.

An interesting feature of these CMDs is the slight excess luminosity of the late-type members with respect to the model isochrone. This is best observed 
in the \gaia\ CMD (right panel in Fig.~\ref{fig:isoc}), because it includes the faintest and coolest members. 
This is reminiscent of what has been observed for M dwarfs in $\gamma$~Velorum \citep{Jeffries2017} and Pleiades \citep{Jackson2018} clusters. They ascribe 
this excess to stellar radii inflation driven by the strong magnetic fields of these young stars that inhibit convection and produce dark spots.

The three earliest members shown in the CMDs are S\,866 (=\,HD\,216057), S\,466 and S\,362 (=\,HD\,214956). The first one is a well-known Be star with an estimated 
spectral type B5Vne \citep{Catanzaro2013a} while the other two objects are A0V stars \citep[spectral types according to this work and][respectively]{Kharchenko2005}. 
The presence of these early-type members suggests an age for the cluster slightly greater than 100 Myr, which is in good agreement with the best-fitting isochrones shown in 
Fig.\,\ref{fig:isoc}.

\section{Stellar parameters and abundances}
\label{Sec:anal}

\subsection{Radial velocity}
\label{subsec:RV}

The first step of our spectroscopic analysis is the measure of the radial velocity, which, as already mentioned in Sect.~\ref{Sec:Observations},
has been obtained by means of the cross-correlation of the HARPS-N spectra of our targets with synthetic templates. For doing this we used the 
task {\sc fxcor} of the IRAF\footnote{IRAF is distributed by the National Optical Astronomy Observatory, which is operated by the Association of 
the Universities for Research in Astronomy, inc. (AURA) under cooperative agreement with the National Science Foundation.} package. 
The RV of the single-lined objects and of the components of SB2 systems has been evaluated as the centroid of the CCF peak(s).  
The RVs for the single-lined members are reported in Table~\ref{Tab:Spectra_param} and range from $-7.5$ to $-4.3$\,\kms, if we exclude 
the hottest, fast-rotating star S\,466. We remark that the errors are rather large, especially for the rapidly rotating stars. The median value of RV is 
$-5.6$\,\kms, while the weighted average is $-4.5$\,\kms.
The values of RV for the components of SB2 systems are reported in Table~\ref{Tab:Spectra_param_SB2}.

\subsection{Atmospheric parameters}
\label{subsec:APS}

We have adopted the code \ROTFIT\ \citep{Frasca2006} to derive the basic atmospheric parameters (APs) for the stars with single-lined HARPS-N
spectra (either single stars or SB1 systems). 
This code allows us to measure also the projected rotational velocity (\vsini) and to perform an MK spectral classification. 
\ROTFIT\  uses a grid of template spectra and is based on a $\chi^2$ minimization of the difference \textit{observed--template} in selected spectral 
regions. 
The grid of templates is composed of high-resolution spectra of real stars with well-known APs, which were retrieved from the ELODIE archive. 
This grid is the same as that adopted in the \gaia-ESO Survey by the Catania (OACT) analysis node \citep[see, e.g.,][]{Smiljanic2014,Frasca2015}. 
When using this grid, the HARPS-N spectra of the targets are degraded to the resolution of ELODIE ($R=42\,000$), which is fully suitable for our 
purposes, also because of the large rotational velocities. 
The target spectra have been also resampled on the ELODIE spectral points ($\Delta\lambda=0.05$\,\AA). 
This has also the advantage to improve the SNR of the spectra to be analyzed.

Each template is aligned in wavelength with the target spectrum thanks to the CCF in a suitable spectral region.
After being aligned with the target spectrum, each template is broadened by convolution with a rotational profile of increasing \vsini\ until a 
minimum $\chi^2$ is attained. The weighted average of the parameters of the best 10 templates is taken for each of the analyzed regions. 

We analyzed 28 spectral segments of 100 \AA\ each from 4000\,\AA\  to 6800\,\AA, which is the range covered by the ELODIE templates and 
corresponds to the portion of HARPS-N spectra where the SNR is the best. The core of Balmer lines, as well as regions severely affected by 
telluric absorption have been disregarded.
The final parameters have been obtained by averaging the results of the individual segments, weighting them according to the $\chi^2$ and the 
amount of information contained in each segment, which is evaluated as the total line absorption, $f_i=\int(F_{\lambda}/F_{\rm C}-1)d\lambda$, 
where $F_{\lambda}/F_{\rm C}$ is the continuum-normalized spectrum in the $i$-th spectral segment.

\begin{figure}
\begin{center}
\hspace{-1.0cm}
\includegraphics[width=9.8cm]{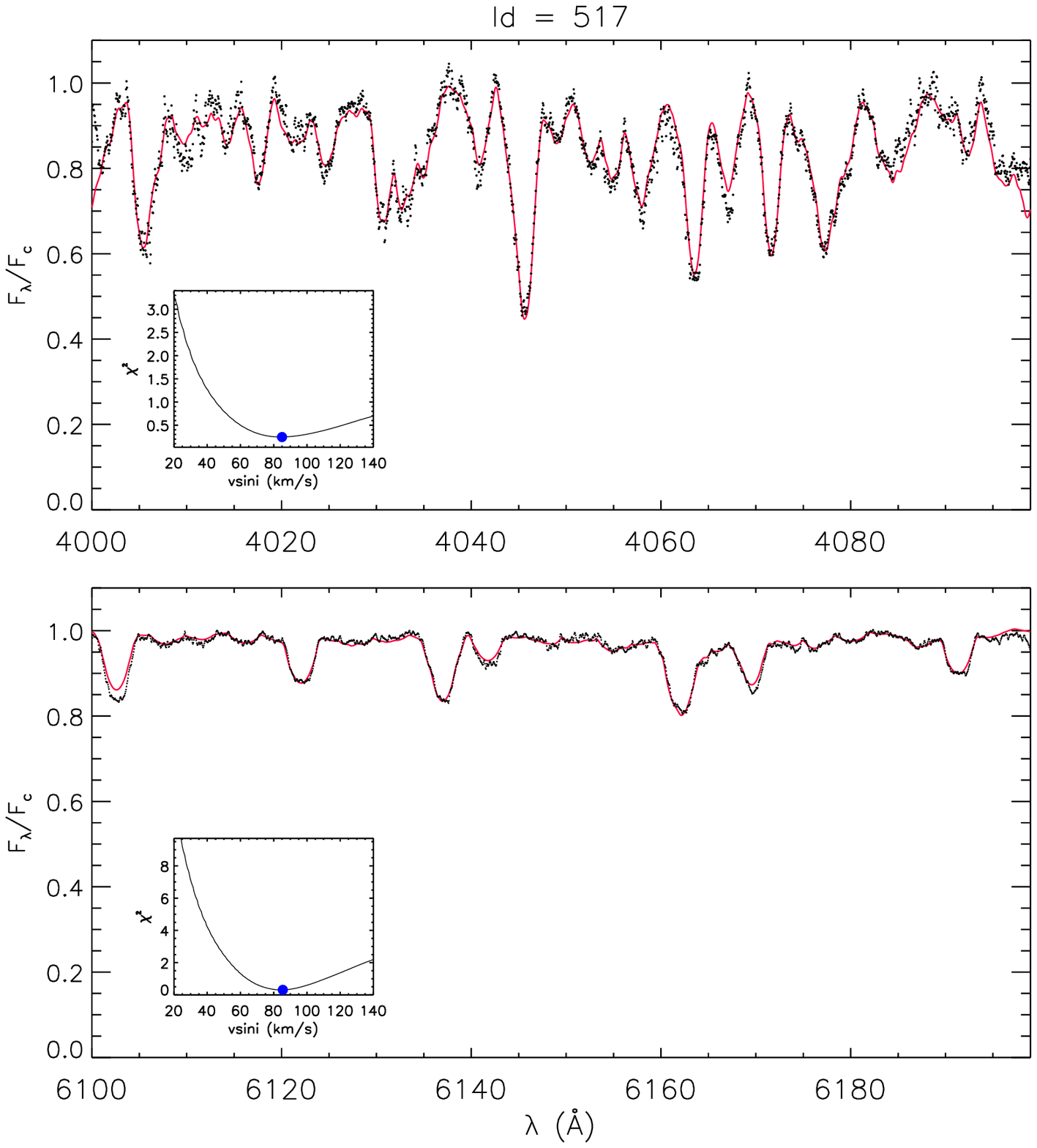}
\vspace{-3.5cm}
\caption{Observed HARPS-N spectrum of S\,517 (black dots) in the $\lambda$\,4000\,\AA\ ({\it upper panel}) and $\lambda$\,6100\,\AA\
({\it lower panel}) spectral regions. In each panel the template spectrum broadened at the \vsini\ of the target is overplotted with a full red line. 
The insets show the $\chi^2$ of the fit as a function of the \vsini. }
\label{fig:spe_517}
\end{center}
\end{figure}

An example of the application of the code \ROTFIT\ to two spectral segments of S\,517 is 
shown in Fig.~\ref{fig:spe_517}, where the observed and template spectra are displayed with black dots and a red line, respectively. 
In the same figure, the $\chi^2$ as a function of \vsini\ is plotted in the insets. 

The results of the analysis for the spectra of single/SB1 systems are summarized in Table~\ref{Tab:Spectra_param}.

\setlength{\tabcolsep}{2pt}
\setlength{\tabcolsep}{4pt}

\begin{table*}[htb]
\caption{Stellar parameters of the single/SB1 systems derived with the code \ROTFIT.}
\begin{tabular}{lrrrrcrrrrc}
\hline
\hline
\noalign{\smallskip}
Id       	   & \teff & err & \logg & err & SpT & $RV$ & err & \vsini & err & $A_V^a$   \\  
           	   & \multicolumn{2}{c}{(K)} & \multicolumn{2}{c}{(dex)} & & \multicolumn{2}{c}{(\kms)}  & \multicolumn{2}{c}{(\kms)}& (mag) \\  
\hline
\noalign{\smallskip}
  39 &  	6667 &  115 & 4.18 & 0.13  & F4V    &	$-$6.36 & 3.31 &  49.1 &  1.9  &  0.01 \\ 
 214 &  	5804 &   87 & 4.41 & 0.12  & G1.5V  &   $-$6.39 & 3.92 & 100.9 &  3.0  &  0.23 \\ 
 378 &  	7636 &  293 & 4.18 & 0.14  & A7III  &   $-$5.6  & 10.9 &  71.1 &  5.8  &  0.18 \\ 
 435 &  	5758 &   79 & 4.39 & 0.12  & G2.5V  &	$-$4.41 & 0.26 &  11.6 &  0.7  &  0.14 \\ 
 466 &         10500 &  250 & 4.00 & 0.20  & A0V    &   $-$13.6 & 5.5  & 260.0 & 15.0  &  0.15 \\ 
{\it 490}$^*$ & 6128 &  113 & 3.95 & 0.11  & F8V    &   $-$15.80& 0.20 &   2.9 &  1.8  &  0.06 \\ 
 517 &  	5784 &   81 & 4.39 & 0.12  & G2V    &   $-$7.03 & 5.59 &  83.6 &  1.9  &  0.36 \\ 
 554 &  	6871 &  152 & 4.14 & 0.13  & F4V    &	$-$5.21 & 6.41 &  81.8 &  3.3  &  0.20 \\ 
{\it 565}$^*$ & 6683 &  112 & 4.02 & 0.12  & F4V    &	$-$1.85 & 2.45 &  23.8 &  0.8  &  0.15 \\ 
 731 &          7355 &  274 & 4.00 & 0.20  & A7III  &   $-$7.54 & 8.41 &  68.9 &  7.0  &  0.07 \\ 
 F1  &          5263 &   92 & 4.55 & 0.10  & K0V    &   $-$4.35 & 0.13 &   6.6 &  0.6  &  0.00 \\	       
 F2  &          5237 &   77 & 4.54 & 0.10  & K1V    &   $-$4.54 & 0.12 &   7.5 &  0.6  &  0.03 \\
\hline
\end{tabular}
\begin{list}{}{}
\item[$^a$] Derived from the SED fitting. 
\item[$^*$] Non member.
\end{list}
\label{Tab:Spectra_param}
\end{table*}

\begin{figure}
\begin{center}
\hspace{-1.0cm}
\includegraphics[width=9.8cm]{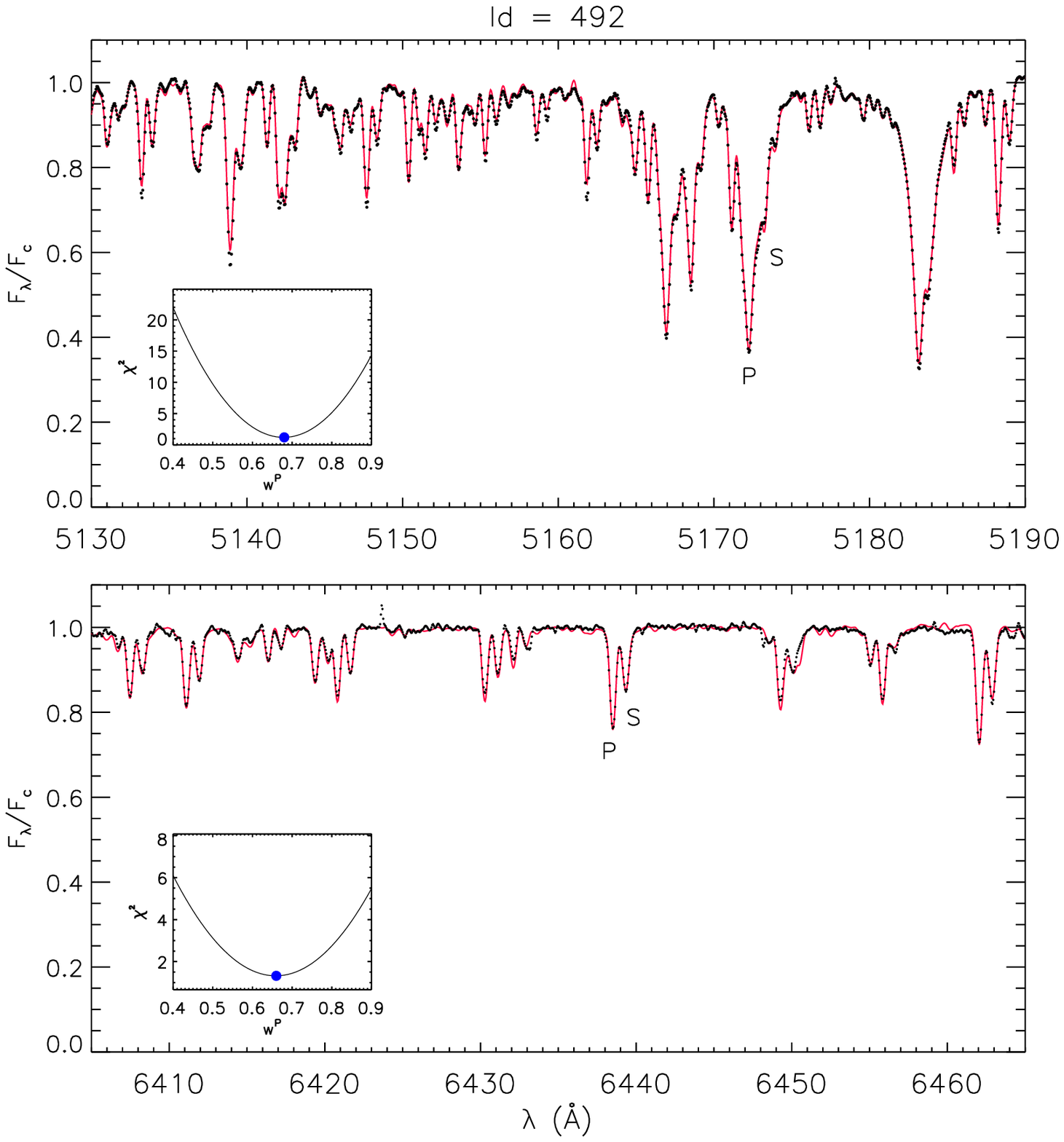}
\vspace{-3.5cm}
\caption{Observed HARPS-N spectrum of the SB2 system S\,492 (black dots) around the \ion{Mg}{i}\,b lines ($\lambda$\,5160\,\AA, {\it upper panel}) and around 
$\lambda$\,6440\,\AA\ ({\it lower panel}). In each panel the ``synthetic'' spectrum, which is the weighted sum of two standard star spectra mimicking
the primary and secondary component of S\,492, is overplotted with a full red line. The insets show the $\chi^2$ of the fit as a function of the contribution
of the primary spectrum, $w^{\rm P}$. 
}
\label{fig:spe_492}
\end{center}
\end{figure}

Among the 17 observed targets we found five spectroscopic double-lined (SB2) systems. For four of them the spectral lines (and the peaks of the CCFs) 
are sufficiently separated in wavelength to allow us to meaningfully infer the stellar parameters of the components of these systems from the analysis of 
their spectra.
To this aim we used \COMPO, a code developed in {\sf IDL}\footnote{IDL (Interactive Data Language) is a registered trademark of  Harris Corporation.} 
environment by \citet{Frasca2006}, which, like \ROTFIT, was adapted to the HARPS-N spectra.
\COMPO\ works in a similar way as \ROTFIT, i.e. it adopts a grid of templates (ELODIE spectra of real stars) to reproduce the observed composite spectrum, 
which is split in 28 segments of 100 \AA\ each that are independently analyzed. 
In this case, the projected rotation velocities of the two components, \vsini$_1$ and \vsini$_2$, are not free parameters, but are computed from the 
full width at half maximum (FWHM) of the peaks of the CCF. The RV separation of the two components is derived from the centroids of the two 
CCF peaks. 
The continuum flux ratio (i.e. the flux contribution of the primary component in units 
of the continuum, $w^{\rm P}$) is  an adjustable parameter. An example of the application of \COMPO\ is shown in Fig.~\ref{fig:spe_492} for two 
spectral segments, one around the \ion{Mg}{i}\,b triplet and the other centered at 6440\,\AA, in which the lines of the two components are clearly distinguishable.

\begin{figure}
\begin{center}
\hspace{-.3cm}
\includegraphics[width=9.5cm]{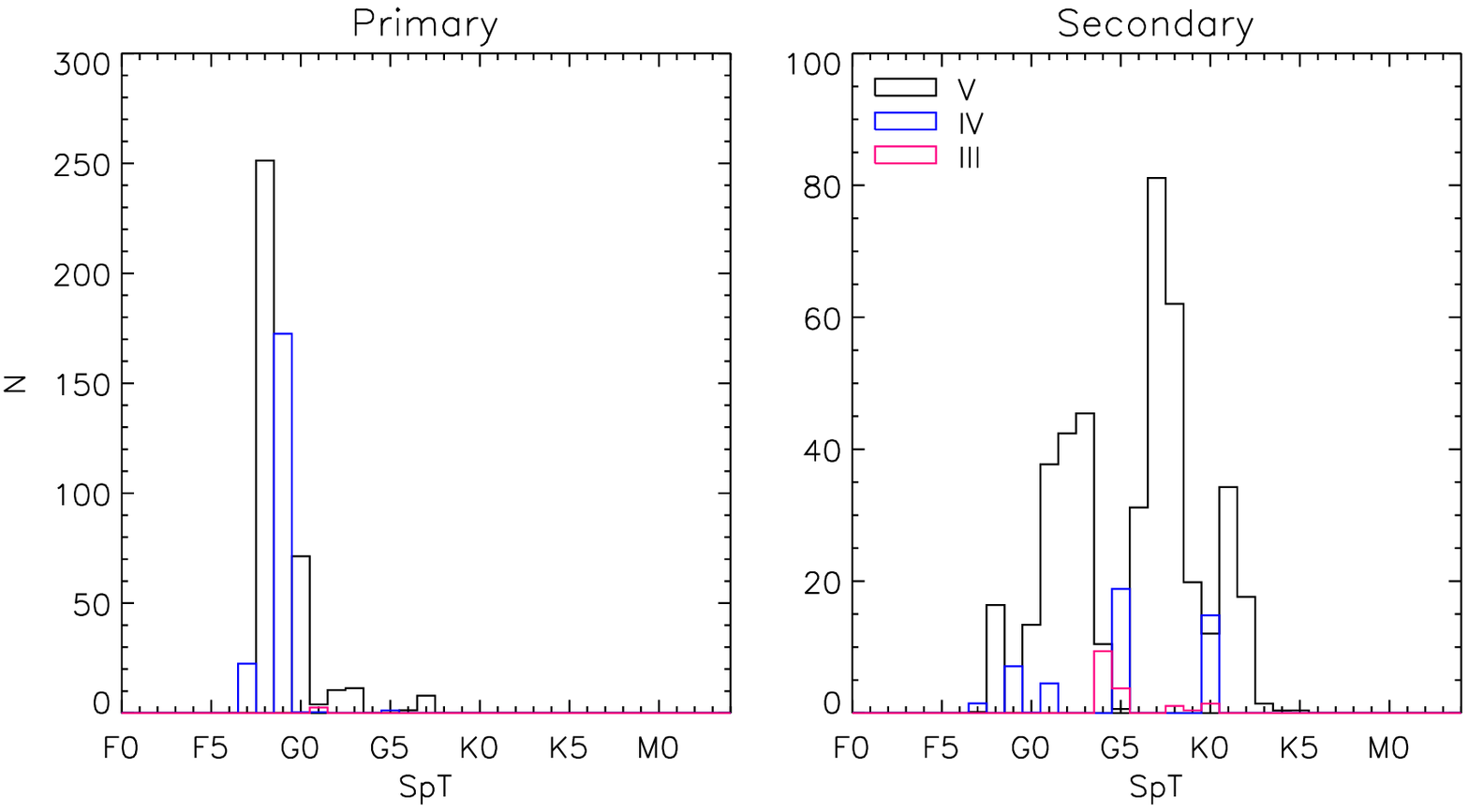}
\vspace{-0.2cm}
\caption{Distribution of spectral types for the components of S\,492.}
\label{fig:histo}
\end{center}
\end{figure}

To evaluate the APs we kept only the best 100 combinations (in terms of minimum $\chi^2$) of primary and secondary spectra per each spectral segment. 
The results for individual segments have been weighted as explained above for \ROTFIT\ to calculate the average APs, which are 
listed in Table~\ref{Tab:Spectra_param_SB2}. The flux contribution has been evaluated at three wavelengths (4400\,\AA, 5500\,\AA, and 6400\,\AA) using 
only the spectral segments around these wavelengths. The relative contribution of the two components changes appreciably from blue to red wavelengths 
only if they have a very different \teff. The spectral types (SpT) of the components are taken as the mode of the spectral-type distributions 
(see Fig.~\ref{fig:histo} for an example).

The system S\,56 displays a complex spectral and CCF behavior, which is reminiscent of an SB2 with two fast-rotating stars 
(\vsini$_1\sim$\,50\,\kms, \vsini$_2\sim$\,70\,\kms) with blended lines in both spectra. We tried a determination of the stellar parameters of the two 
components with \COMPO\ and obtained a decent fit of the observed spectrum in several spectral regions. However, the stellar parameters, 
especially those of the fainter, faster-rotating secondary, must be considered as indicative only.

\setlength{\tabcolsep}{3pt}

\begin{table*}[htb]
\caption{Stellar parameters of the components of SB2 systems derived with the code \COMPO.}
\begin{tabular}{cccccccccccccc}
\hline
\hline
\noalign{\smallskip}
\multirow{3}{*}{Id}       	   & \teff & err & \logg &  err  & $RV$ & err & $\gamma$  & \multirow{3}{*}{$w^{\rm P}_{4400}$} &  \multirow{3}{*}{$w^{\rm P}_{5500}$} & \multirow{3}{*}{$w^{\rm P}_{6400}$} & SpT & \vsini  \\ 
           	   &  (K)  & (K) & (dex) & (dex) &  \multicolumn{2}{c}{(\kms)} &  (\kms) &                  &        	     &  		  &         & (\kms)  \\  
           	   & [P/S] &[P/S]& [P/S] & [P/S] &  [P/S] & [P/S] &   & &                &  		  &  [P/S]  & [P/S]  \\  
\hline
\noalign{\smallskip}
56$^a$&{\scriptsize 6670/6420} & {\scriptsize 150/250} & {\scriptsize 4.30/4.20} & {\scriptsize 0.15/0.20} & {\scriptsize 13.3/-56.3} & {\scriptsize 4.5/13.3}& {\scriptsize -18.5$\pm$6.8} & {\scriptsize 0.68$\pm$0.12}& {\scriptsize 0.66$\pm$0.10} & {\scriptsize \dots}         & {\scriptsize F2V/F7IV}  & {\scriptsize 50/70} \\ 
 266 & {\scriptsize 6094/5550} & {\scriptsize  64/209} & {\scriptsize 4.15/4.26} & {\scriptsize 0.11/0.22} & {\scriptsize -78.3/54.7} & {\scriptsize 1.4/0.6} & {\scriptsize -20.2$\pm$1.3} & {\scriptsize 0.79$\pm$0.03} & {\scriptsize 0.75$\pm$0.03} & {\scriptsize 0.74$\pm$0.04} & {\scriptsize F8V/G3V}	& {\scriptsize 10/5 } \\  
 492 & {\scriptsize 6075/5560} & {\scriptsize  59/168} & {\scriptsize 4.24/4.32} & {\scriptsize 0.06/0.17} & {\scriptsize -26.2/12.9} & {\scriptsize 0.9/1.8} & {\scriptsize  -8.2$\pm$1.0} & {\scriptsize 0.69$\pm$0.03} & {\scriptsize 0.67$\pm$0.02} & {\scriptsize 0.65$\pm$0.04} & {\scriptsize F9V/G7V}	& {\scriptsize 12/12} \\  
{\it 502}$^*$ & {\scriptsize 5730/5628} & {\scriptsize 111/152} & {\scriptsize 4.32/4.36} & {\scriptsize 0.11/0.13} & {\scriptsize -19.7/10.8} & {\scriptsize 0.5/0.6} & {\scriptsize  -4.7$\pm$0.4} & {\scriptsize 0.55$\pm$0.02} & {\scriptsize 0.54$\pm$0.02} & {\scriptsize 0.53$\pm$0.03} & {\scriptsize G2V/G2V}	& {\scriptsize  5/5 } \\  
 708 & {\scriptsize 6114/6099} & {\scriptsize  81/183} & {\scriptsize 4.12/4.18} & {\scriptsize 0.16/0.19} & {\scriptsize -25.7/34.6} & {\scriptsize 0.7/5.6} & {\scriptsize   4.0$\pm$2.8} & {\scriptsize 0.54$\pm$0.06} & {\scriptsize 0.53$\pm$0.05} & {\scriptsize 0.50$\pm$0.05} & {\scriptsize F8IV/F8IV} & {\scriptsize 10/50} \\  
\hline
\end{tabular}
\begin{list}{}{}
\item[$^a$] Uncertain parameters due to the high \vsini\ values and the strong line blend.
\item[$^*$] Non member.
\end{list}
\label{Tab:Spectra_param_SB2}
\end{table*}

For the SB2 systems we have also evaluated the systemic velocity $\gamma$. In an ideal case of two components with the same mass, $\gamma$ is the 
mean of the radial velocities, $\gamma=(RV_1+RV_2)/2$. In general, the systemic velocity can be calculated as

\begin{equation}
\gamma = RV_1 + \frac{(RV_2 - RV_1)}{\frac{M_1}{M_2}+1},
\label{Eq:gamma}
\end{equation}

\noindent{where $RV_1$ and $RV_2$ are the radial velocities and $M_1$ and $M_2$ the masses of the two components.}

As we have only one or two spectra per each system, we cannot derive the mass ratio, $M_1/M_2$, which can only be derived by the solution of a full RV curve. 
Therefore, $M_1/M_2$ must be estimated by the spectra of these systems.
Assuming that all the components are on the main sequence (which is probably not valid for S\,56), we can use a mass-luminosity relation for estimating
the mass ratio.
To this purpose we have adopted the relation proposed by \citet[][table 3]{Eker2015} for intermediate-mass stars, $\log L = 4.328\times\log M -0.002$,
from which we get

\begin{equation}
\frac{M_1}{M_2} = \left(\frac{L_1}{L_2}\right)^{0.23}.
\label{Eq:mass_ratio}
\end{equation}

\noindent{The luminosity ratio $\frac{L_1}{L_2}$ has been approximated by the ratio of flux contributions at 5500\,\AA\ (reported in 
Table~\ref{Tab:Spectra_param_SB2}). The error of $\gamma$ has been calculated propagating the errors of $RV_1$, $RV_2$, and $w^{\rm P}_{5500}$ through 
Eqs.~\ref{Eq:gamma} and \ref{Eq:mass_ratio}.
The values of $\gamma$ are also reported in Table~\ref{Tab:Spectra_param_SB2} along with their errors.}

The $\gamma$ values span a range similar to that of the RVs for the single-lined objects, with the exception, perhaps, of S\,708 and S\,266.
However, repeated observations are needed to get more accurate values of $\gamma$.

\subsection{Elemental abundances}
\label{subsec:abundance}

We have also analyzed the HARPS-N spectra of the single-lined targets by means of a spectral synthesis approach 
\citep[e.g.,][]{Catanzaro2011,Catanzaro2013}. For this task we did not degrade the spectra but maintained 
their original resolution $R=115\,000$. 
We used the {\sf ATLAS9} code \citep{Kurucz1993a,Kurucz1993b} to compute LTE atmospheric models and {\sf SYNTHE}
\citep{Kurucz1981} to produce synthetic spectra that include both the instrumental and rotational broadening.

As a first step, we used the APs obtained with \rotfit, but we made a check letting them vary and finding the best values by  
minimizing the $\chi^2$ of the difference \textit{observed--synthetic}. We always found values consistent with those of \rotfit\ 
within the errors. We have therefore adopted the \rotfit\ atmospheric parameters reported in Table~\ref{Tab:Spectra_param} and used 
them for the subsequent analysis.

We have then derived the abundances of elements of atomic number up to 40.
In particular, we analyzed independently thirty-nine 50-\AA-wide spectral segments between 4400 and 6800\,\AA.
Ad-hoc routines written in {\sf IDL} allowed us to find the best solution by $\chi^2$ minimization. 
The elemental abundances, expressed in the standard notation $A($X$)= \log\,$[n(X)/n(H)]\,+\,12, are listed in 
Table~\ref{Tab:abundances}.    
The weighted-averaged abundances for the cluster (considering only the confirmed members) are reported in Table~\ref{Tab:abund_cluster_sun} along
with the standard error of the weighted mean. 
In this table we also report the abundances derived by us for the Sun, applying the same procedure to a HARPS-N spectrum of Ganymede acquired 
in the framework of the GAPS program \citep[Global Architecture of Planetary Systems,][]{Covino2013}. 

\setlength{\tabcolsep}{5pt}

\begin{table}
\caption{Average elemental abundances for ASCC\,123 and for the Sun derived in the present work. The values are expressed according to the 
standard notation $A($X$)= \log\,$[n(X)/n(H)]\,+\,12.} 
\centering
\begin{tabular}{lcc}
\hline
\hline
\noalign{\smallskip}
Element         & ASCC\,123  &  Sun$^a$ \\ 
\noalign{\smallskip}
\hline
\noalign{\smallskip}
C  & 8.41\,$\pm$\,0.06  &  8.54\,$\pm$\,0.04  \\
O  & 9.09\,$\pm$\,0.20  &  8.80\,$\pm$\,0.01  \\
Na & 6.13\,$\pm$\,0.08  &  6.17\,$\pm$\,0.14  \\
Mg & 7.71\,$\pm$\,0.04  &  7.70\,$\pm$\,0.10  \\
Al & 6.47\,$\pm$\,0.06  &  6.55\,$\pm$\,0.15  \\
Si & 7.52\,$\pm$\,0.08  &  7.50\,$\pm$\,0.08  \\
S  & 7.42\,$\pm$\,0.05  &  7.33\,$\pm$\,0.10  \\
Ca & 6.60\,$\pm$\,0.04  &  6.32\,$\pm$\,0.09  \\
Sc & 3.52\,$\pm$\,0.04  &  3.14\,$\pm$\,0.13  \\
Ti & 4.99\,$\pm$\,0.07  &  4.86\,$\pm$\,0.12  \\
V  & 4.50\,$\pm$\,0.14  &  3.96\,$\pm$\,0.10  \\
Cr & 5.77\,$\pm$\,0.04  &  5.57\,$\pm$\,0.15  \\
Mn & 5.56\,$\pm$\,0.05  &  5.35\,$\pm$\,0.15  \\
Fe & 7.59\,$\pm$\,0.04  &  7.45\,$\pm$\,0.08  \\
Co & 5.35\,$\pm$\,0.18  &  4.93\,$\pm$\,0.05  \\
Ni & 6.29\,$\pm$\,0.06  &  6.16\,$\pm$\,0.10  \\
Cu & 4.44\,$\pm$\,0.10  &  4.26\,$\pm$\,0.10  \\
Zn & 4.44\,$\pm$\,0.08  &  4.53\,$\pm$\,0.09  \\
Sr & 3.55\,$\pm$\,0.08  &  2.95\,$\pm$\,0.15  \\
Y  & 2.66\,$\pm$\,0.16  &  2.25\,$\pm$\,0.04  \\
Zr & 3.16\,$\pm$\,0.08  &  2.65\,$\pm$\,0.14  \\
\hline
\end{tabular}
\begin{list}{}{}
\item[$^a$] Derived from a HARPS-N spectrum of Ganymede.
\end{list}
\label{Tab:abund_cluster_sun}
\end{table}

\begin{figure}
\begin{center}
\hspace{-0.5cm}
\includegraphics[width=9.cm]{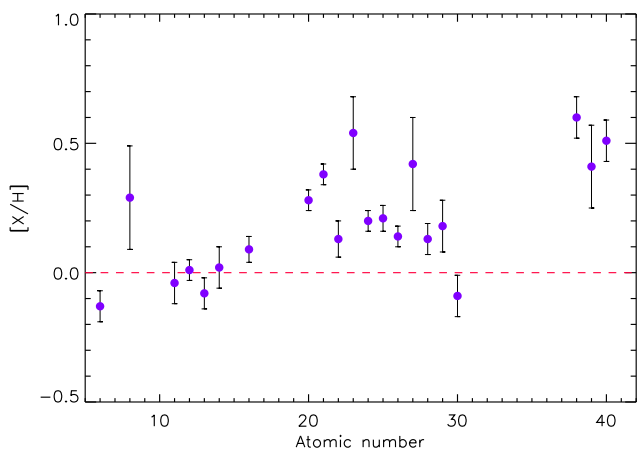}
\vspace{0.cm}
\caption{Chemical pattern for ASCC\,123. The horizontal dashed line corresponds to the solar abundances reported 
in Table~\ref{Tab:abund_cluster_sun}. } 
\label{fig:abundances}
\end{center}
\end{figure}

The pattern of chemical abundances for ASCC\,123 is shown in Fig.~\ref{fig:abundances}, where we have used as reference the 
solar abundances derived in the present work (Table~\ref{Tab:abund_cluster_sun}). We note an average overabundance of $\sim\,0.2$\,dex 
with respect to solar values for the iron-peak elements.  
In particular, the iron abundance for the cluster is [Fe/H]=+0.14\,$\pm$\,0.04. The heavy elements (Sr, Y, and Zr) are overabundant with respect to
the Sun by about 0.5\,dex.

The chemical pattern for the individual stars in the field of ASCC\,123 is displayed in Fig.~\ref{fig:abundances_individual}.
As apparent from this figure, the non-member S\,490 shows all the elements underabundant compared to the Sun, with an average value of 
$\sim -0.3$\,dex. The other likely non-member S\,565 displays a solar chemical abundance.

These results are discussed in more detail in Sect.~\ref{Sec:disc}.

\subsection{Spectral energy distributions and Hertzsprung-Russell diagram}
\label{subsec:HR}

For the single stars, we used the spectral energy distribution (SED) to evaluate the interstellar extinction, $A_V$, and to 
derive the luminosities. The SEDs were built with optical and near-infrared (NIR) photometric data available 
in the literature and reported in Table~\ref{Phot_123}. 

The SEDs were fitted with NextGen low-resolution synthetic spectra \citep{Hauschildt1999}. 
For each target, we adopted the {\it Gaia} DR2 parallax and kept \teff\ and \logg\  fixed to the values derived with \rotfit\ 
(Table~\ref{Tab:Spectra_param}), while the stellar radius, $R_*$, and $A_V$ were set as free parameters.
The best values of $R_*$ and $A_V$ were found by  $\chi^2$ minimization. 
The luminosity has been obtained as $4\pi R_*^2\sigma T_{\rm eff}^4$.

We found values of $A_V$ in the range 0.00--0.36\,mag (see Table~\ref{Tab:Spectra_param}) with  
an average of 0.13\,mag. Assuming a typical value $R_V=3.1$ for the total-to-selective extinction, we obtain a reddening 
$E(B-V)\simeq 0.04$\,mag, which is lower than the value of 0.15\,mag reported by \citet{Kharchenko2013} for this cluster.

We have derived the luminosities of the components of the SB2 systems from the combined $V$ magnitude listed in Table~\ref{Phot_123} 
and the luminosity ratio at 5500\,\AA\ stemming from  $w^{\rm P}_{5500}$ (Table~\ref{Tab:Spectra_param_SB2}). The values of $V$ have been
corrected for the extinction ($A_V=0.13$\,mag) and have been used to calculate the absolute magnitudes with the \gaia\ distances. The bolometric 
correction of  \citet{PecautMamajek2013} was applied and the bolometric magnitude of the Sun, $M_{\rm bol}^{\sun}=4.64$\,mag \citep{Cox2000}, 
was used to express the stellar luminosity in solar units.
We expect that the luminosities of the components of SB2 systems are less accurate than for the single stars as a result of the more complex 
way to extract the information from the data. 

\begin{figure}
\begin{center}
\hspace{-.7cm}
\includegraphics[width=9.5cm]{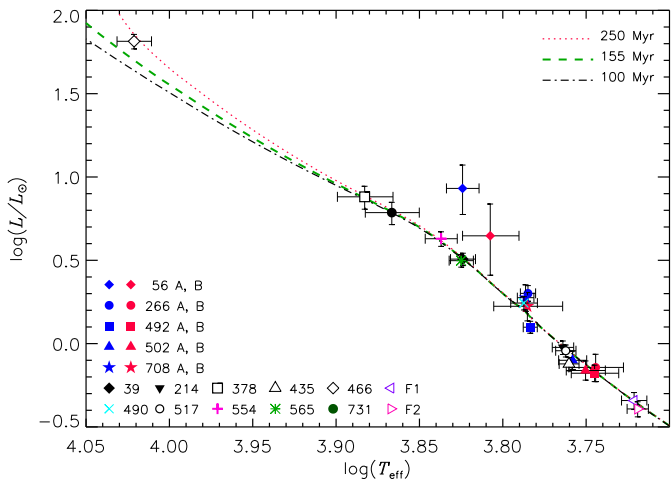}
\vspace{-0.2cm}
\caption{HR diagram. The stars are coded with different symbols and colors. PARSEC isochrones at 100, 155, and 250\, Myr for models 
with Z=0.021 ([Fe/H]\,=\,+0.14) are overlaid. 
}
\label{fig:HR}
\end{center}
\end{figure}

The Hertzsprung-Russell (HR) diagram of our targets is displayed in Fig.~\ref{fig:HR} where we overplot PARSEC isochrones 
at Z\,=\,0.021 ([Fe/H]\,=\,+0.14, see Sect.\,\ref{subsec:abundance}) at three ages.
As apparent from this figure, most of the targets are located in the region of the HR diagram where the MS isochrones 
overlap, so that very little information about the cluster age can be gathered from this plot.

The position of the two components of S\,56 is significantly above the aforementioned isochrones, suggesting that these stars are more 
evolved (older age) or that some mass exchange has occurred between the components of the system. However, as outlined in Sec.~\ref{subsec:APS}, 
the parameters of the components of this system are very uncertain and should be taken with caution. 

The position of the hottest source in our sample, S\,466, is in between the 155-Myr and 250-Myr isochrones.
This is in line with the CMDs of Fig.~\ref{fig:isoc} in which it (upper red dot) is slightly displaced redward with respect to the 155-Myr isochrone. 
However, the hottest member of the cluster, HD~216057, seems to fit well on the 155-Myr isochrone in CMDs (upper gray dot in each panel of 
Fig.~\ref{fig:isoc}).

\section{Chromospheric emission and lithium abundance}
\label{Sec:chrom_lithium}

\begin{figure}
\begin{center}
\hspace{-.5cm}
\includegraphics[width=9.0cm]{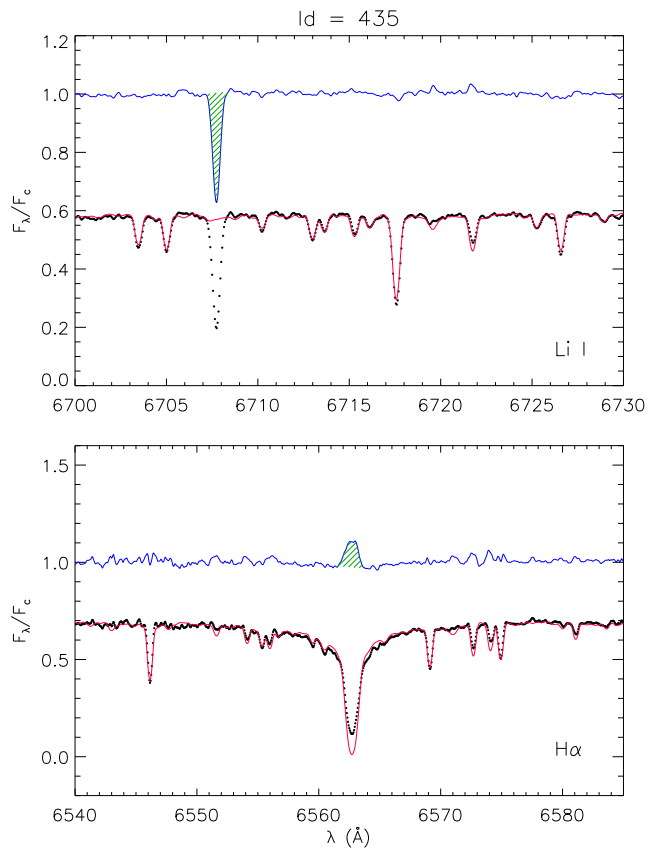}
\vspace{-1.5cm}
\caption{Subtraction of the non-active, lithium poor template (red line) from
the spectrum of S\,435 (black dots), which reveals the chromospheric emission in the H$\alpha$ core (blue
line in the {\it bottom panel}) and emphasizes the \ion{Li}{i} $\lambda$6708\,\AA\ absorption line,
removing the nearby blended lines ({\it top panel}). The green hatched areas represent the excess H$\alpha$ 
emission ({\it bottom panel}) and \ion{Li}{i} absorption ({\it top panel}) that have been integrated to
obtain $W_{\rm H\alpha}^{em}$ and $W_{\rm Li}$, respectively.}
\label{fig:subtraction}
\end{center}
\end{figure}

For stars with an age from a few ten to a few hundred Myr, the H$\alpha$ emission (diagnostic of chromospheric activity) and lithium 
absorption can be used to estimate the age \citep[see, e.g.,][and references therein]{Jeffries2014, Frasca2018}.

To this aim, the \textit{synthetic} spectra produced by \ROTFIT\ and \COMPO\ with non-active, lithium-poor templates have been subtracted to 
the observed spectra of the targets to measure the excess emission in the core of the H$\alpha$ line ($EW_{\rm H\alpha}^{em}$) and the equivalent 
width of the \ion{Li}{i} $\lambda$6708\,\AA\ absorption line ($EW_{\rm Li}$), removing the blends with nearby lines.

\setlength{\tabcolsep}{3pt}

\begin{table}[htb]
\caption{H$\alpha$, \ion{Li}{i}$\lambda$6708\,\AA\ equivalent widths and lithium abundance for the single/SB1 systems and the components of SB2 systems.}
\begin{center}
\begin{tabular}{lrrrcrrlr}
\hline
\hline
\noalign{\smallskip}
Id       	   & \teff &  $EW_{\rm H\alpha}^{em}$    & err & $R'_{\rm H\alpha}$  & $EW_{\rm Li}$ & err & $A$(Li) & err   \\  
           	   & (K)   &  \multicolumn{2}{c}{(m\AA)}       &                     &  \multicolumn{2}{c}{(m\AA)}  & \multicolumn{2}{c}{(dex)}   \\  
\hline
\noalign{\smallskip}
  39    &  6667 &   73  &   13  & $-5.09^{+0.07}_{-0.09}$  &  59 &  8 & {\it 2.97}$^a$  &  0.08  \\ 
 214    &  5804 &  291  &   21  & $-4.46^{+0.04}_{-0.04}$  & 211 & 14 & 3.42  &  0.19  \\ 
 435    &  5758 &  144  &   18  & $-4.76^{+0.06}_{-0.06}$  & 193 &  6 & 3.24  &  0.12  \\ 
 490    &  6128 &   24  &   11  & $-5.55^{+0.17}_{-0.27}$  &  7  &  3 & 1.72  &  0.33  \\ 
 517    &  5784 &  591  &   33  & $-4.15^{+0.03}_{-0.03}$  & 223 & 16 & 3.47  &  0.20  \\
 554    &  6871 & \dots & \dots & \dots  &  60 &  9 & {\it 2.98}$^a$  &  0.08  \\
 565    &  6683 & \dots & \dots & \dots  &  63 &  4 & {\it 3.00}$^a$  &  0.04  \\
  56\,P &  6670 & \dots & \dots & \dots  &  85 & 15 & {\it 3.18}$^a$  &  0.11  \\ 
 266\,P &  6094 & \dots & \dots & \dots  & 106 &  4 & 3.00  &  0.08  \\ 
 266\,S &  5550 & \dots & \dots & \dots  & 182 & 12 & 2.95  &  0.30  \\ 
 492\,P &  6075 &136$^b$&   11  & \dots  & 111 &  6 & 3.01  &  0.09  \\ 
 492\,S &  5558 & \dots & \dots & \dots  &  73 & 12 & 2.26  &  0.28  \\ 
 502\,P &  5730 & 64$^b$&    9  & \dots  &  70 & 11 & 2.41  &  0.20  \\ 
 502\,S &  5628 & \dots & \dots & \dots  &  65 & 13 & 2.27  &  0.26  \\ 
 708\,P &  6114 & \dots & \dots & \dots  & 174 &  7 & 3.45  &  0.12  \\ 
 708\,S &  6099 & \dots & \dots & \dots  &  58 & 25 & 2.64  &  0.39  \\ 
  F1    &  5263 &  196  &   22  & $-4.63^{+0.05}_{-0.06}$  & 163  &  9 & 2.51 & 0.16 \\
  F2    &  5237 &  396  &   30  & $-4.32^{+0.04}_{-0.05}$  & 213  & 10 & 2.79 & 0.16 \\
\hline
\end{tabular}
\end{center}
\begin{list}{}{}
\item[$^a$] \teff$>$6500\,K. $A$(Li) extrapolated from the \citet{Soderblom1993c} tables. 
\item[$^b$] Blended H$\alpha$ residual emission profile. The total residual emission has been integrated.
\end{list}
\label{Tab:Halpha_Lithium}
\end{table}

\begin{figure}
\begin{center}
\hspace{-.5cm}
\includegraphics[width=9.0cm]{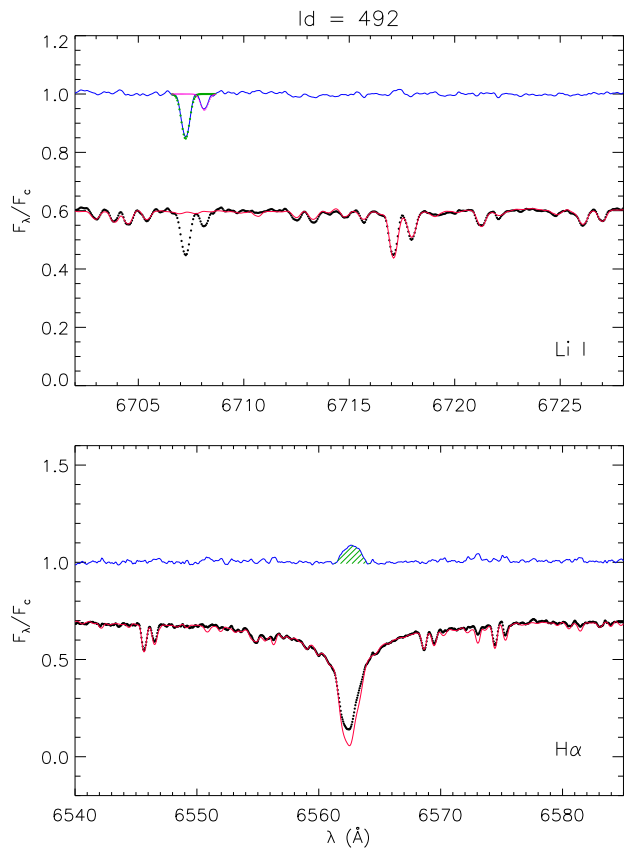}
\vspace{-1.5cm}
\caption{Subtraction of the `synthetic' composite spectrum (red line) from the observed spectrum of 
the SB2 S\,492 (black dots), which leaves only the chromospheric emission in the H$\alpha$ core (blue line in the {\it bottom panel}) and emphasizes 
the \ion{Li}{i} $\lambda$6708\,\AA\ absorption lines of the two component (blue line in the {\it top panel}). 
The green and magenta dotted lines superimposed to the residual spectrum represent Gaussian fits used to deblend 
the \ion{Li}{i} lines of the two components.
The two components cannot be separated in the H$\alpha$ and the full residual H$\alpha$ emission can be integrated to provide a
total $W_{\rm H\alpha}^{em}$ (green hatched area in the {\it bottom panel}).}
\label{fig:subtraction_compo2}
\end{center}
\end{figure}

An example of spectral subtraction for a single-lined spectrum is shown in Fig.~\ref{fig:subtraction}.
$EW_{\rm H\alpha}^{em}$ and $EW_{\rm Li}$ have been measured on the ``subtracted'' spectra (blue lines in Fig.~\ref{fig:subtraction}) by 
integrating the residual emission and absorption profiles, respectively, and are quoted in Table~\ref{Tab:Halpha_Lithium}. 
For SB2 systems, subtraction is even more necessary to remove different lines of the two components which, due to the different Doppler 
shift, overlap one another. The difference spectrum contains only the \ion{Li}{i} absorption lines of the two components, whose equivalent 
widths can eventually be measured individually by means of a two-Gaussian deblending (see Fig.~\ref{fig:subtraction_compo2} for an example). 
We point out that the values of $EW_{\rm Li}$ for the components of SB2 systems, which are quoted in Table~\ref{Tab:Halpha_Lithium}, have been 
corrected for the contribution to the continuum of the respective component.   

Lithium is a fragile element that is burned in stellar interiors at temperature as low as 2.5$\times10^6$\,K. It is progressively depleted 
from the stellar atmosphere in a way depending on the internal structure (i.e. stellar mass) for stars with convective envelopes deep enough to 
reach temperatures at which lithium is burned. 
It can therefore be used as an age proxy for stars cooler than about 6500\,K. 

\begin{figure}
\begin{center}
\hspace{-.5cm}
\includegraphics[width=9.0cm]{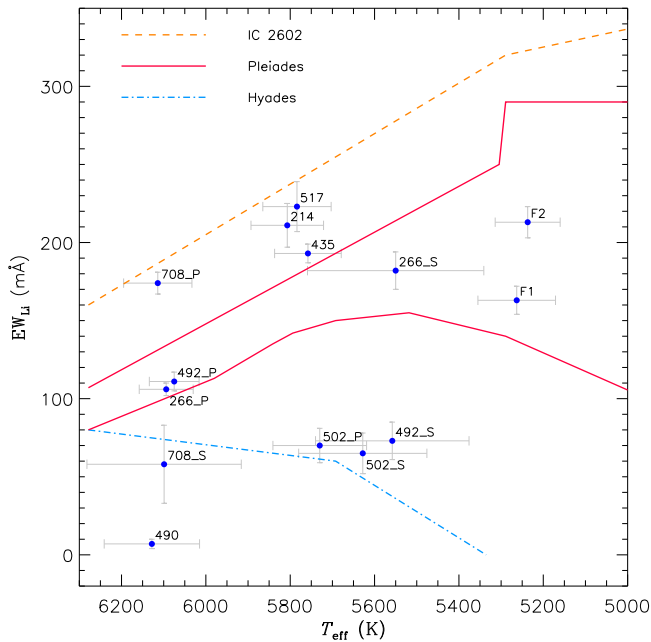}
\vspace{0cm}
\caption{Lithium equivalent width versus effective temperature for the single stars and the components of SB2 systems. 
The upper envelopes (and lower envelope for the Pleiades) of a few young clusters are overplotted. The points are labeled with the
star Id, as listed in Table~\ref{Tab:Halpha_Lithium}. }
\label{fig:lithium_ew}
\end{center}
\end{figure}

A simple way to get an estimate of stellar ages is a diagram which shows $EW_{\rm Li}$ or lithium abundance, $A$(Li), versus \teff\ for the investigated 
stars and to compare their position with the upper envelopes of clusters with a known age. 
The isochronal age estimated for ASCC\,123 in the literature range from about 130\,Myr \citep{Yen2018} to 260\,Myr \citep{Kharchenko2005}.
The CMDs of the present paper (Fig.~\ref{fig:isoc}) do not allow us to notably improve the age determination, which is $\tau\approx$\,150\,Myr with an uncertainty
of at least 50\,Myr. This is basically due to the lack of turn-off cluster members, as above mentioned. 
For this reason and to identify anomalous/contaminant stars in our sample, we used as comparison the upper envelopes of the following clusters: the Hyades 
\citep{Soderblom1990}, the Pleiades \citep{Soderblom1993c,Neuhauser1997}, and IC~2602 \citep{Montes2001}, whose ages are of about 650~Myr, 
125~Myr \citep{White2007AJ133}, and 30~Myr \citep{Stauffer1997}, respectively. 
As shown by \citet{Soderblom1993b}, the lower envelope of the Pleiades behaves as an upper boundary for the members of the Ursa Major (UMa) cluster 
($\tau\approx$\,300 Myr).

The \teff--$EW_{\rm Li}$ diagram is shown in Fig.~\ref{fig:lithium_ew}. According to the range of isochronal ages, the $EW_{\rm Li}$ values
should be comprised between the lower and upper envelope of the Pleiades cluster.  

We note that S\,490 has a too low $EW_{\rm Li}$, which is totally inconsistent 
with the cluster age. This fact reinforces the hypothesis that this star does not belong to ASCC\,123, as already suggested by its kinematics. 
Both components of the binary system S\,502 display also low $EW_{\rm Li}$ values that are compatible with an older age ($\tau\approx 600$\,Myr).
For this reason, we do not consider S\,502 as a bona-fide member of the cluster.
The other stars/systems display instead higher $EW_{\rm Li}$ values, with the exception of the secondary components of S\,708 and S\,492.
However, the primary components of these two systems show $EW_{\rm Li}$ values compatible with a young age. We remark that $EW$ measurement is 
a more difficult task for binary system components, since the contribution to the continuum also plays a role. This could affect the weaker components more. 
Nevertheless, we cannot exclude that binarity may have an effect in the depletion of lithium. 

We note that four objects lie above the upper envelope of the Pleiades and below that of IC\,2602, suggesting an age younger than 100\,Myr
or an age spread. However, the two early K-type members display a lithium content in very good agreement with the Pleiades. 
To answer these issues, high- or mid-resolution spectra of lower mass members, which are too faint for GIARPS, would be of great help.

We derived $A$(Li) from $EW_{\rm Li}$, \teff\ and \logg\  by interpolating the curves of growth of \citet{Soderblom1993c}, which have been preferred to 
those of \citet{Pavlenko1996} because the former encompass the \teff\ range 4000--6500\,K, while the latter, although calculated in NLTE, end at 6000\,K.
However, the differences of $A$(Li) for the stars with \teff\,$\leq 6000$\,K are at most 0.15 dex. The values of $A$(Li) are reported in 
Table~\ref{Tab:Halpha_Lithium}.

\begin{figure}[htb]  
\begin{center}
\includegraphics[width=8.8cm]{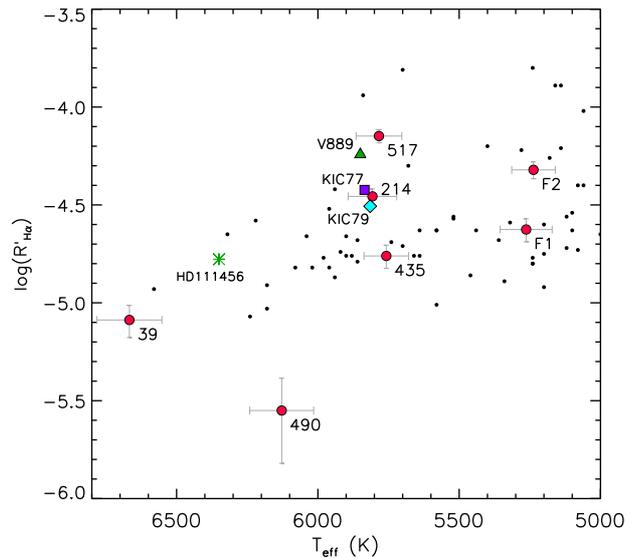}  
\vspace{-.2cm}
\caption{$R'_{\rm H\alpha}$ as a function of $T_{\rm eff}$ for the single-lined objects (red dots). 
The stars are labeled according to their Id. The small black dots denote the measures made by \citet{Soderblom1993a} for Pleiades stars. The triangle refers to 
V889~Her \citep{Frasca2010}, while the diamond and square denote the two Sun-like stars KIC\,7985370 and KIC\,7765135, respectively \citep{Froehlich2012}. 
The asterisk refers to the young F-type star HD\,111546 \citep{Freire2004}.}
\label{Fig:Flux_Teff}
 \end{center}
\end{figure}

As regards the H$\alpha$, we detected a relevant filling in the line core for all the three G-type stars.
Two of them, S\,214 and S\,517, are ultrafast rotators, with \vsini=80--100\,\kms, while S\,435 is rotating slower (\vsini$\simeq 12$\,\kms)
but still displays a H$\alpha$ core filling. 

For these stars we computed the H$\alpha$ surface flux, $F_{\rm H\alpha}$, and the ratio of the H$\alpha$ and bolometric luminosity, 
$R'_{\rm H\alpha}$ as:
\begin{eqnarray}
F_{\rm H\alpha} & = & F_{6563}W_{\rm H\alpha}^{em}, 
\end{eqnarray}
\begin{eqnarray}
R'_{\rm H\alpha}&  = & L_{\rm H\alpha}/L_{\rm bol} = F_{\rm H\alpha}/(\sigma T_{\rm eff}^4),
\end{eqnarray}
{\noindent where $F_{6563}$ is the flux at the continuum near the H$\alpha$ line per unit stellar surface area, which has been evaluated from the 
NextGen synthetic low-resolution spectra \citep{Hauschildt1999} at the stellar temperature and surface gravity of the target. We evaluated the flux error by taking 
into account the error of $W_{\rm H\alpha}^{em}$ and the uncertainty in the continuum flux at the line center, $F_{6563}$, which is estimated considering the errors of \teff\  
and \logg. }

$R'_{\rm H\alpha}$ is plotted against \teff\ in Fig.~\ref{Fig:Flux_Teff} along with the data for the Pleiades stars from \citet{Soderblom1993a}, the 
UMa-cluster member HD\,111456 \citep{Freire2004}, and three young Sun-like stars in the field, namely V889~Her \citep[$\tau \approx50$\,Myr,][]{Frasca2010}, 
KIC\,7985370 and KIC\,7765135 \citep[$\tau=100-200$\,Myr,][]{Froehlich2012}. 
The chromospheric emission of the members of ASCC\,123 is comparable with that of the Pleiades and young Sun-like stars, confirming that the former are indeed young stars 
($\tau=50-200$\,Myr) but cannot provide us with absolute ages. 
However, it is worth to note that for the three Sun-like targets $R'_{\rm H\alpha}$ and $EW_{\rm Li}$ are correlated, i.e. more active is the star, higher is the lithium
abundance.

\section{Discussion}
\label{Sec:disc}

\begin{figure}  
  \centering 
  \includegraphics[width=\columnwidth]{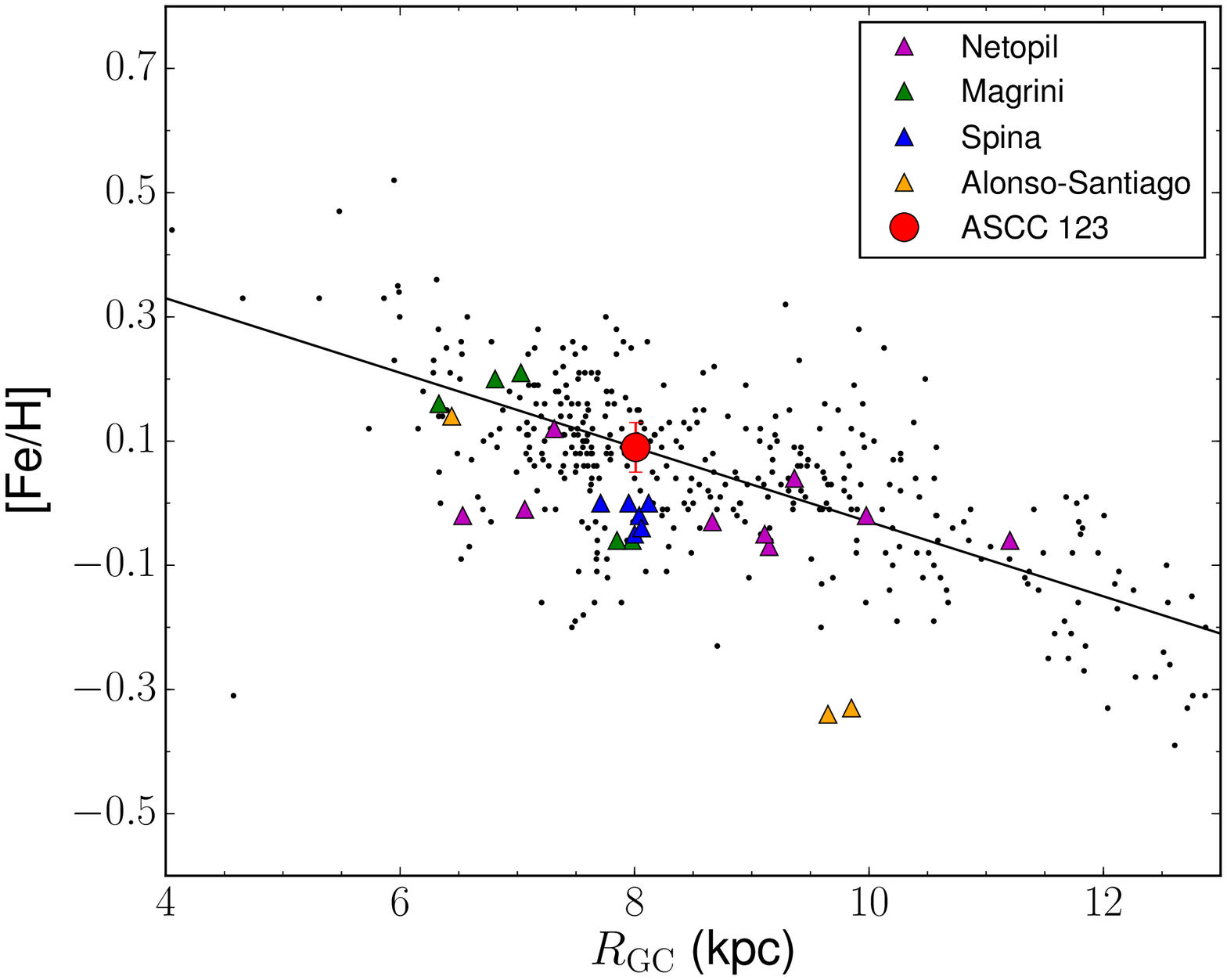} 
  \caption{Iron abundance gradient in the Milky Way found by \citet{Genovali13,Genovali14} based on Cepheids (small black dots).
  The black line is the Galactic gradient ($-$0.06~dex/kpc).
 Triangles represent young open clusters ($<500\:$Myr) studied by \citet{Netopil2016}, \citet{Spina17}, \citet{Magrini18}, and \citet{6067,3105,2345}. 
 Finally, the red circle is ASCC\,123.
  All the values shown in this plot are rescaled to \citet{Genovali14}, i.e. R$_{\sun}$=7.95 kpc and A(Fe)=7.50.} 
  \label{gradient_ascc} 
\end{figure}

We performed a comprehensive study of the cluster ASCC\,123 resorting to different approaches and datasets, such as \gaia\ astrometry, photometry and 
high-resolution spectroscopy. All together these data allow us to better define the properties of this cluster.

The atmospheric parameters, particularly \teff, along with the photometry allowed us to evaluate the average extinction towards the cluster 
as $A_V\simeq 0.13$\,mag, i.e. \ebv\,$\simeq 0.04$\,mag with $R_V=3.1$, which is a little smaller than the value of \ebv\,$\simeq$\,0.10\,mag reported 
in the literature for this cluster \citep[e.g.,][]{Kharchenko2001,Yen2018}. However, it is in agreement with the Gaia-2MASS 3D extinction maps of 
\citet[][see their figure A.1 at $l=105\degr$]{Lallement2019}. 

We have used different diagnostics to constrain the age of the cluster, namely the CMDs, the HR diagram, the chromospheric activity level and the lithium 
abundance. The determination of extinction and metallicity has allowed us to select the proper isochrones to fit the CMDs and HR diagram. Unfortunately, 
the absence of turn-off stars prevents us from an accurate determination of the cluster age from the isochrones. However, an upper limit of 
$\tau\approx 250$\,Myr can be defined. 
The lithium abundance and the H$\alpha$ flux suggest an age similar to that of the Pleiades ($\tau\simeq 125$\,Myr) or a little younger, but certainly not 
older than the UMa cluster ($\tau\simeq 300$\,Myr). Therefore, we can only infer an age range of 100--250\,Myr. 

As already commented in Sect.\,\ref{Sec:intro}, this is the first time that a chemical analysis is performed on ASCC\,123. In the absence of any previous study with 
which to do a comparative discussion of our results we resorted to the Galactic gradient as a reference for a proper comparison. \citet{Genovali13,Genovali14} traced 
the distribution of the metallicity in the Galactic disk, in terms of [Fe/H], as a function of the Galactocentric distance. For this task they used Cepheids, with ages 
comprised between 20 and 400 Myr, which makes them very suitable to be compared to young open clusters. 
In Fig.\,\ref{gradient_ascc} ASCC\,123 is overplotted on this gradient. Other young clusters ($\tau\leq$\,500\,Myr) taken from \citet{Netopil2016},
\citet{Spina17}, \citet{Magrini18}, and \citet{6067,3105,2345} are added for further comparison. We note that ASCC\,123  (with an average [Fe/H]=+0.14$\pm$0.04) 
follows closely the average trend with the Galactocentric distance.  This result might seem at odds with previous findings of a solar or sub-solar metallicity
for young and intermediate-age clusters \citep[see, e.g.,][and references therein]{Dorazi2011,Spina17}. However, it is in agreement with what is expected from standard 
chemical evolution (e.g., \citealt{Minchev2013}, and references therein) and recent results of a super-solar metallicity ([Fe/H]\,$\approx$\,+0.2) for some young or
intermediate-age clusters in the solar neighborhood, such as NGC\,6067 \citep[][]{6067} and Praesepe \citep[][]{Dorazi2019}.

We also compare our abundances with the Galactic trends found in the solar neighborhood by the \gaia-ESO Survey for the thin and thick discs 
(data release 4)\footnote{\tt https://www.eso.org/qi/catalogQuery/index/121}. 
 In Fig.~\ref{trends1_ascc} and \ref{trends2_ascc} we plot some abundance ratios [X/Fe] vs [Fe/H]. 
The chemical composition of ASCC\,123 is generally in agreement with the Galactic trends observed in the thin disc of the Galaxy. 
A few elements (V, Co, and Y) seem to be slightly overabundant, but still compatible with the Galactic trends within the errors. 
Instead, Na and Zn appear slightly underabundant, but the discrepancy from the bulk of \gaia-ESO Survey data, at least for sodium, 
is $\sim$ 1\,sigma and therefore it is not significant. The underabundance of Zn could be marginally significant.

\section{Summary}
\label{Sec:summary}

In the framework of the SPA project, we have studied a sample of 17 candidate members to the young open cluster ASCC\,123. 
We have performed, for the first time, a thorough analysis of their properties based on high-resolution spectroscopy with the aim of improving our 
knowledge of this nearby and very sparse cluster. 

We have measured their radial and projected rotational velocities finding five new SB2 systems.
We have derived their atmospheric parameters, chromospheric activity level and lithium abundance by means of our own tools. For the single stars 
we also carried out a detailed chemical analysis.

We found that three out of the 17 targets have to be considered as non-members on the basis of their astrometry and lithium content.

From the likely single-lined members, which are all MS stars, we derived an average extinction $A_V=0.13$\,mag, a median RV$=-5.6$\,\kms, and 
an average iron abundance [Fe/H]$=+0.14$, which is in very good agreement with the Galactic metallicity gradient.
We also infer a chemical composition of the cluster compatible with the solar neighborhood, as traced by the \gaia-ESO survey.
For ASCC\,123 only few elements display abundance ratios slightly discrepant with respect to these average trends. Those with the largest differences 
are Zn (underabundant) and Y (overabundant). 

The analysis of lithium content (for the stars with \teff$<6500$\,K) shows that six targets have an abundance compatible or even higher than that
measured for the Pleiades, a nearly coeval cluster. 
The chromospheric activity level, as traced by the index $R'_{\rm H\alpha}$, for these stars is also in line with that of the Pleiades and of young 
Sun-like stars.
This data confirm the young age of ASCC\,123 inferred from the color-magnitude diagrams, suggesting a range of about 100--250 Myr.

\begin{acknowledgements}
We thank the referee for very useful comments and suggestions.
We acknowledge the support from the Italian {\it Ministero dell'Istruzione, Universit\`a e  Ricerca} (MIUR).
This research made use of SIMBAD and VIZIER databases, operated at the CDS, Strasbourg, France. 
This work has made use of data from the European Space Agency (ESA)
mission {\it Gaia} ({\tt https://www.cosmos.esa.int/gaia}), processed by
the {\it Gaia} Data Processing and Analysis Consortium (DPAC,
{\tt https://www.cosmos.esa.int/web/gaia/dpac/consortium}). Funding
for the DPAC has been provided by national institutions, in particular
the institutions participating in the {\it Gaia} Multilateral Agreement.
We made use of data from the \gaia-ESO Survey Data Archive, prepared and hosted by 
the Wide Field Astronomy Unit, Institute for Astronomy, University of Edinburgh, 
which is funded by the UKScience and Technology Facilities Council.
This publication made use of data products from the Two Micron All Sky Survey, 
which is a joint project of the University of Massachusetts and the Infrared Processing and 
Analysis Center/California Institute of Technology, funded by the National Aeronautics and Space 
Administration and the National Science Foundation

\end{acknowledgements}

\bibliographystyle{aa}

{}

\newpage

\appendix

\section{Additional tables and figures}

\begin{landscape}
\begin{table}
\caption{Photometric data for the stars observed in this work.}
\begin{center}
\begin{tabular}{lccccccccc}  
\hline\hline
Id & $V^a$ & $(B-V)^a$ &  $I^b$  &  $J$ & $H$ & $K_{\textrm{s}}$ & $G$ & $G_{\rm BP}$ & $G_{\rm RP}$ \\
\hline
\noalign{\smallskip}
\multicolumn{9}{c}{Members}\\
\noalign{\smallskip}
\hline
\noalign{\smallskip}
39  & 10.433 $\pm$ 0.040 & 0.567 $\pm$ 0.185 &  9.964 $\pm$ 0.069 &  9.558 $\pm$ 0.018 &  9.369 $\pm$ 0.017 &  9.291 $\pm$ 0.017  & 10.3523 $\pm$ 0.0003 & 10.6014 $\pm$ 0.0006 & 9.9768 $\pm$ 0.0007 \\
56  &  9.140 $\pm$ 0.019$^c$ & 0.426 $\pm$ 0.026$^c$ & 8.728 $\pm$ 0.064  &  8.289 $\pm$ 0.023 &  8.118 $\pm$ 0.018 &  8.084 $\pm$ 0.017  &  9.0516 $\pm$ 0.0003  &  9.2806 $\pm$ 0.0013  &  8.7063 $\pm$ 0.0016 \\
214 & 12.042 $\pm$ 0.044 & 0.654 $\pm$ 0.113 & 11.358 $\pm$ 0.115 & 10.674 $\pm$ 0.020 & 10.289 $\pm$ 0.017 & 10.194 $\pm$ 0.016  & 11.8297 $\pm$ 0.0020  & 12.2201 $\pm$ 0.0065  & 11.2949 $\pm$ 0.0058 \\ 
266 & 10.701 $\pm$ 0.055 & 0.548 $\pm$ 0.089 & \dots &  9.637 $\pm$ 0.021 &  9.384 $\pm$ 0.018 &  9.314 $\pm$ 0.017  & 10.5989 $\pm$ 0.0005  & 10.9141 $\pm$ 0.0009  & 10.1484 $\pm$ 0.0009 \\
378 &  9.610 $\pm$ 0.024$^c$ & 0.299 $\pm$ 0.031$^c$ &  9.273 $\pm$ 0.086 &  8.976 $\pm$ 0.026 &  8.796 $\pm$ 0.021 &  8.714 $\pm$ 0.022  &  9.5476 $\pm$ 0.0003  &  9.7109 $\pm$ 0.0012  &  9.2895 $\pm$ 0.0016 \\
435 & 12.164 $\pm$ 0.048 & 0.853 $\pm$ 0.148 & 11.334 $\pm$ 0.074 & 10.845 $\pm$ 0.025 & 10.457 $\pm$ 0.030 & 10.418 $\pm$ 0.024  & 12.0500 $\pm$ 0.0012  & 12.4520 $\pm$ 0.0044  & 11.5002 $\pm$ 0.0032 \\
466 &  7.580 $\pm$ 0.010$^c$ & 0.034 $\pm$ 0.010$^c$ &  7.474 $\pm$ 0.072 &  7.387 $\pm$ 0.030 &  7.392 $\pm$ 0.044 &  7.396 $\pm$ 0.018  &  7.5500 $\pm$ 0.0003  &  7.5910 $\pm$ 0.0013  &  7.5102 $\pm$ 0.0020 \\
492 & 10.786 $\pm$ 0.041 & 0.656 $\pm$ 0.077 & 10.100 $\pm$ 0.059 &  9.543 $\pm$ 0.027 &  9.234 $\pm$ 0.033 &  9.168 $\pm$ 0.018  & 10.6354 $\pm$ 0.0009  & 11.0087 $\pm$ 0.0029  & 10.1256 $\pm$ 0.0018 \\
517 & 12.201 $\pm$ 0.060 & 0.742 $\pm$ 0.088 & 11.410 $\pm$ 0.079 & 10.709 $\pm$ 0.022 & 10.300 $\pm$ 0.031 & 10.205 $\pm$ 0.019  & 12.0120 $\pm$ 0.0017  & 12.4433 $\pm$ 0.0063  & 11.4328 $\pm$ 0.0050 \\
554 & 10.271 $\pm$ 0.038 & 0.415 $\pm$ 0.066 &  9.757 $\pm$ 0.079 &  9.364 $\pm$ 0.022 &  9.190 $\pm$ 0.030 &  9.168 $\pm$ 0.025  & 10.1712 $\pm$ 0.0004  & 10.4320 $\pm$ 0.0009  &  9.7893 $\pm$ 0.0010 \\
708 & 10.152 $\pm$ 0.038 & 0.538 $\pm$ 0.060 &  9.578 $\pm$ 0.085 &  9.041 $\pm$ 0.027 &  8.824 $\pm$ 0.030 &  8.742 $\pm$ 0.021  & 10.0149 $\pm$ 0.0006  & 10.3262 $\pm$ 0.0016  &  9.5728 $\pm$ 0.0013 \\
731 &  9.700 $\pm$ 0.032$^c$ & 0.316 $\pm$ 0.041$^c$ &  9.325 $\pm$ 0.063 &  8.973 $\pm$ 0.035 &  8.885 $\pm$ 0.028 &  8.837 $\pm$ 0.020  &  9.5734 $\pm$ 0.0005  &  9.7564 $\pm$ 0.0016  &  9.3056 $\pm$ 0.0018 \\
 F1 & 12.732 $\pm$ 0.035 & 0.803 $\pm$ 0.056 & 11.68 $\pm$ 0.22$^d$ &  11.138 $\pm$ 0.020 & 10.761 $\pm$ 0.015 & 10.649 $\pm$ 0.023 & 12.5669 $\pm$ 0.0008 & 13.0236 $\pm$ 0.0033 & 11.9585 $\pm$ 0.0024 \\
 F2 & 12.961 $\pm$ 0.040 & 0.867 $\pm$ 0.049 & 12.07 $\pm$ 0.43$^d$ &  11.365 $\pm$ 0.020 & 10.954 $\pm$ 0.017 & 10.839 $\pm$ 0.020 & 12.7646 $\pm$ 0.0013 & 13.2418 $\pm$ 0.0034 & 12.1487 $\pm$ 0.0025 \\
\hline
\noalign{\smallskip}
\multicolumn{9}{c}{Non members}\\
\noalign{\smallskip}
\hline
\noalign{\smallskip}
490 & 11.504 $\pm$ 0.030 & 0.536 $\pm$ 0.079 & 10.972 $\pm$ 0.078 & 10.366 $\pm$ 0.019 & 10.148 $\pm$ 0.026 & 10.075 $\pm$ 0.022  & 11.3770 $\pm$ 0.0007  & 11.6955 $\pm$ 0.0012  & 10.9154 $\pm$ 0.0007 \\
502 & 11.180 $\pm$ 0.034 & 0.745 $\pm$ 0.072 & 10.455 $\pm$ 0.097 &  9.846 $\pm$ 0.030 &  9.476 $\pm$ 0.036 &  9.429 $\pm$ 0.024  & 11.0200 $\pm$ 0.0008  & 11.4338 $\pm$ 0.0016  & 10.4729 $\pm$ 0.0012 \\
565 & 10.454 $\pm$ 0.060 & 0.451 $\pm$ 0.077 &  9.953 $\pm$ 0.060 &  9.458 $\pm$ 0.021 &  9.252 $\pm$ 0.016 &  9.180 $\pm$ 0.015  & 10.3391 $\pm$ 0.0003  & 10.6116 $\pm$ 0.0006  &  9.9356 $\pm$ 0.0008 \\
\hline
\end{tabular}
\begin{list}{}{}
\item[$^a$] From the APASS catalog \citep{APASS}.
\item[$^b$] From the TASS catalog \citep{TASS}.
\item[$^c$] From the TYCHO catalog \citep{HIPPA97}.
\item[$^d$] $N$ photographic magnitude from GSC2.3 catalog \citep{Lasker2008}.
\end{list}{}{}

\label{Phot_123}
\end{center}
\end{table}
\end{landscape}

\begin{landscape}
\begin{table}
\caption{Elemental abundances for the single/SB1 stars expressed according to the standard notation $A($X$)= \log\,$[n(X)/n(H)]\,+\,12.}  
\begin{tabular}{lcccccccccccc}
\hline
\hline
\noalign{\smallskip}
Element  & 39          & 214               & 378               & 435               & 466               & {\it 490}$^*$     & 517               &  554              & {\it 565}$^*$     & 731               &   F1   &   F2  \\ 
\noalign{\smallskip}
\hline
\noalign{\smallskip}
\hline
\noalign{\smallskip}
C  & \dots             & \dots             & 8.49\,$\pm$\,0.15 & \dots             & 8.22\,$\pm$\,0.13 & 8.07\,$\pm$\,0.12 & \dots             & 8.44\,$\pm$\,0.12 & 8.37\,$\pm$\,0.14 & 8.50\,$\pm$\,0.12 & \dots             & \dots  	   \\
O  & \dots             & \dots             & 8.50\,$\pm$\,0.21 & \dots             & 8.81\,$\pm$\,0.17 & \dots             & \dots             & 9.82\,$\pm$\,0.15 & 8.75\,$\pm$\,0.15 & 8.79\,$\pm$\,0.18 & \dots             & \dots  	   \\
Na & 5.99\,$\pm$\,0.29 & 6.05\,$\pm$\,0.10 & 6.06\,$\pm$\,0.19 & 5.70\,$\pm$\,0.18 & \dots             & 5.37\,$\pm$\,0.08 & 5.99\,$\pm$\,0.15 & \dots             & 6.23\,$\pm$\,0.15 & 5.75\,$\pm$\,0.25 & 6.26\,$\pm$\,0.14 & 6.39\,$\pm$\,0.09 \\
Mg & 7.76\,$\pm$\,0.19 & 7.76\,$\pm$\,0.08 & 8.06\,$\pm$\,0.12 & 7.68\,$\pm$\,0.04 & 7.59\,$\pm$\,0.06 & 7.25\,$\pm$\,0.16 & 7.79\,$\pm$\,0.11 & 7.78\,$\pm$\,0.11 & 7.50\,$\pm$\,0.15 & 7.76\,$\pm$\,0.10 & 7.61\,$\pm$\,0.18 & 7.80\,$\pm$\,0.09 \\
Al & \dots             & \dots             & 6.61\,$\pm$\,0.50 & 6.23\,$\pm$\,0.15 & \dots             & 6.14\,$\pm$\,0.10 & \dots             & \dots             & 6.96\,$\pm$\,0.15 & 6.40\,$\pm$\,0.12 & 6.58\,$\pm$\,0.09 & 6.52\,$\pm$\,0.15 \\
Si & 7.59\,$\pm$\,0.18 & 7.18\,$\pm$\,0.14 & 7.79\,$\pm$\,0.18 & 7.35\,$\pm$\,0.10 & 7.27\,$\pm$\,0.17 & 7.00\,$\pm$\,0.18 & 7.34\,$\pm$\,0.20 & 7.52\,$\pm$\,0.10 & 7.40\,$\pm$\,0.19 & 7.82\,$\pm$\,0.13 & 7.68\,$\pm$\,0.10 & 7.78\,$\pm$\,0.19 \\
S  & 7.33\,$\pm$\,0.13 & \dots             & 7.51\,$\pm$\,0.11 & \dots             & \dots             & 7.07\,$\pm$\,0.10 & \dots             & 7.41\,$\pm$\,0.17 & 7.13\,$\pm$\,0.10 & 7.41\,$\pm$\,0.08 & \dots             & \dots  	   \\
Ca & 6.54\,$\pm$\,0.13 & 6.75\,$\pm$\,0.10 & 6.63\,$\pm$\,0.09 & 6.59\,$\pm$\,0.17 & 6.46\,$\pm$\,0.19 & 6.08\,$\pm$\,0.14 & 6.65\,$\pm$\,0.13 & 6.44\,$\pm$\,0.13 & 6.36\,$\pm$\,0.16 & 6.56\,$\pm$\,0.13 & 6.49\,$\pm$\,0.12 & 6.71\,$\pm$\,0.11 \\
Sc & 3.64\,$\pm$\,0.15 & 3.66\,$\pm$\,0.16 & 3.50\,$\pm$\,0.18 & 3.42\,$\pm$\,0.16 & \dots             & 2.67\,$\pm$\,0.10 & 3.63\,$\pm$\,0.15 & 3.67\,$\pm$\,0.10 & 3.07\,$\pm$\,0.14 & 3.50\,$\pm$\,0.13 & 3.41\,$\pm$\,0.08 & 3.39\,$\pm$\,0.12 \\
Ti & 4.95\,$\pm$\,0.17 & 4.96\,$\pm$\,0.18 & 5.14\,$\pm$\,0.12 & 5.00\,$\pm$\,0.16 & 4.26\,$\pm$\,0.19 & 4.57\,$\pm$\,0.12 & 4.95\,$\pm$\,0.16 & 5.09\,$\pm$\,0.11 & 4.86\,$\pm$\,0.11 & 5.00\,$\pm$\,0.10 & 5.01\,$\pm$\,0.10 & 5.10\,$\pm$\,0.13 \\
V  & 4.41\,$\pm$\,0.14 & 4.53\,$\pm$\,0.15 & 4.91\,$\pm$\,0.14 & 4.20\,$\pm$\,0.12 & \dots             & 3.74\,$\pm$\,0.18 & 4.65\,$\pm$\,0.14 & 5.06\,$\pm$\,0.19 & 4.43\,$\pm$\,0.15 & 5.13\,$\pm$\,0.11 & 4.07\,$\pm$\,0.10 & 4.23\,$\pm$\,0.10 \\
Cr & 5.85\,$\pm$\,0.16 & 5.67\,$\pm$\,0.14 & 5.81\,$\pm$\,0.16 & 5.81\,$\pm$\,0.10 & 5.63\,$\pm$\,0.15 & 5.35\,$\pm$\,0.12 & 5.57\,$\pm$\,0.17 & 5.84\,$\pm$\,0.17 & 5.77\,$\pm$\,0.11 & 5.79\,$\pm$\,0.18 & 5.75\,$\pm$\,0.10 & 5.86\,$\pm$\,0.10 \\
Mn & 5.73\,$\pm$\,0.08 & 5.33\,$\pm$\,0.16 & 5.46\,$\pm$\,0.06 & 5.56\,$\pm$\,0.16 & \dots             & 5.09\,$\pm$\,0.11 & 5.57\,$\pm$\,0.14 & 5.69\,$\pm$\,0.19 & 5.45\,$\pm$\,0.13 & 5.57\,$\pm$\,0.19 & 5.59\,$\pm$\,0.12 & 5.67\,$\pm$\,0.14 \\
Fe & 7.50\,$\pm$\,0.14 & 7.68\,$\pm$\,0.15 & 7.76\,$\pm$\,0.18 & 7.61\,$\pm$\,0.12 & 7.69\,$\pm$\,0.11 & 7.03\,$\pm$\,0.11 & 7.53\,$\pm$\,0.11 & 7.62\,$\pm$\,0.09 & 7.34\,$\pm$\,0.11 & 7.61\,$\pm$\,0.17 & 7.47\,$\pm$\,0.13 & 7.57\,$\pm$\,0.11 \\
Co & \dots             & 5.20\,$\pm$\,0.10 & 7.26\,$\pm$\,2.72 & 4.85\,$\pm$\,0.12 & \dots             & 4.78\,$\pm$\,0.12 & 5.34\,$\pm$\,0.16 & 5.34\,$\pm$\,0.17 & 5.45\,$\pm$\,0.18 & 6.09\,$\pm$\,0.08 & 4.93\,$\pm$\,0.10 & 5.03\,$\pm$\,0.12 \\
Ni & 6.27\,$\pm$\,0.10 & 6.17\,$\pm$\,0.17 & 6.39\,$\pm$\,0.12 & 6.24\,$\pm$\,0.15 & 6.69\,$\pm$\,0.11 & 5.88\,$\pm$\,0.11 & 6.07\,$\pm$\,0.09 & 6.32\,$\pm$\,0.15 & 6.19\,$\pm$\,0.16 & 6.26\,$\pm$\,0.10 & 6.27\,$\pm$\,0.16 & 6.27\,$\pm$\,0.11 \\
Cu & \dots             & 4.72\,$\pm$\,0.36 & \dots             & 4.24\,$\pm$\,0.15 & \dots             & 3.96\,$\pm$\,0.11 & 4.36\,$\pm$\,0.15 & 4.81\,$\pm$\,0.11 & \dots             & \dots             & 4.15\,$\pm$\,0.07 & 4.63\,$\pm$\,0.07 \\
Zn & \dots             & \dots             & \dots             & 4.42\,$\pm$\,0.10 & \dots             & 4.12\,$\pm$\,0.10 & 3.95\,$\pm$\,0.18 & \dots             & 4.33\,$\pm$\,0.07 & \dots             & 4.60\,$\pm$\,0.11 & 4.55\,$\pm$\,0.16 \\
Sr & \dots             & \dots             & \dots             & 3.49\,$\pm$\,0.12 & \dots             & \dots             & \dots             & \dots             & 4.26\,$\pm$\,0.35 & \dots             & 3.51\,$\pm$\,0.15 & 3.76\,$\pm$\,0.18 \\
Y  & 2.61\,$\pm$\,0.08 & 3.15\,$\pm$\,0.26 & 2.33\,$\pm$\,0.13 & \dots             & \dots             & 2.10\,$\pm$\,0.15 & 3.78\,$\pm$\,0.15 & 2.51\,$\pm$\,0.16 & 2.50\,$\pm$\,0.15 & 2.25\,$\pm$\,0.12 & 2.72\,$\pm$\,0.17 & 2.82\,$\pm$\,0.18 \\
Zr & 3.30\,$\pm$\,0.14 & 3.19\,$\pm$\,0.10 & 3.66\,$\pm$\,0.19 & 2.82\,$\pm$\,0.08 & \dots             & 2.42\,$\pm$\,0.12 & 3.13\,$\pm$\,0.15 & 3.23\,$\pm$\,0.07 & 3.00\,$\pm$\,0.11 & 3.48\,$\pm$\,0.18 & 3.13\,$\pm$\,0.12 & 3.46\,$\pm$\,0.16 \\
\hline
\end{tabular}
\begin{list}{}{}
\item[$^*$] Non member.
\end{list}
\label{Tab:abundances}
\end{table}
\end{landscape}

\begin{figure*}
\begin{center}
\hspace{-0.3cm}
\includegraphics[width=16.cm]{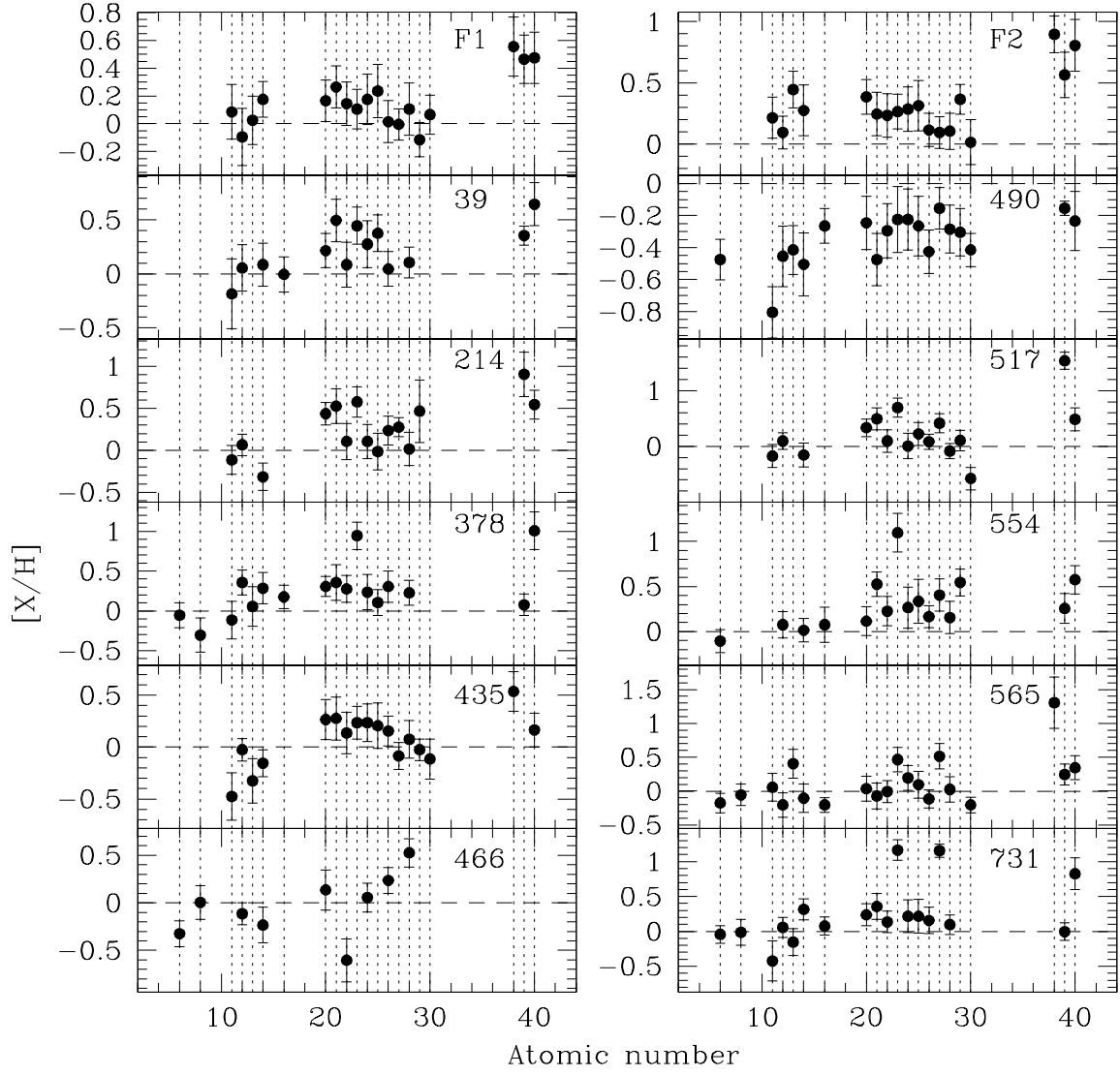}
\vspace{0.cm}
\caption{Chemical patterns derived for our targets. The horizontal dashed line in each box corresponds to the solar abundances reported 
in Table~\ref{Tab:abund_cluster_sun}.} 
\label{fig:abundances_individual}
\end{center}
\end{figure*}

\begin{figure*} 
  \centering         
\vspace{1.cm}
  \includegraphics[width=18cm]{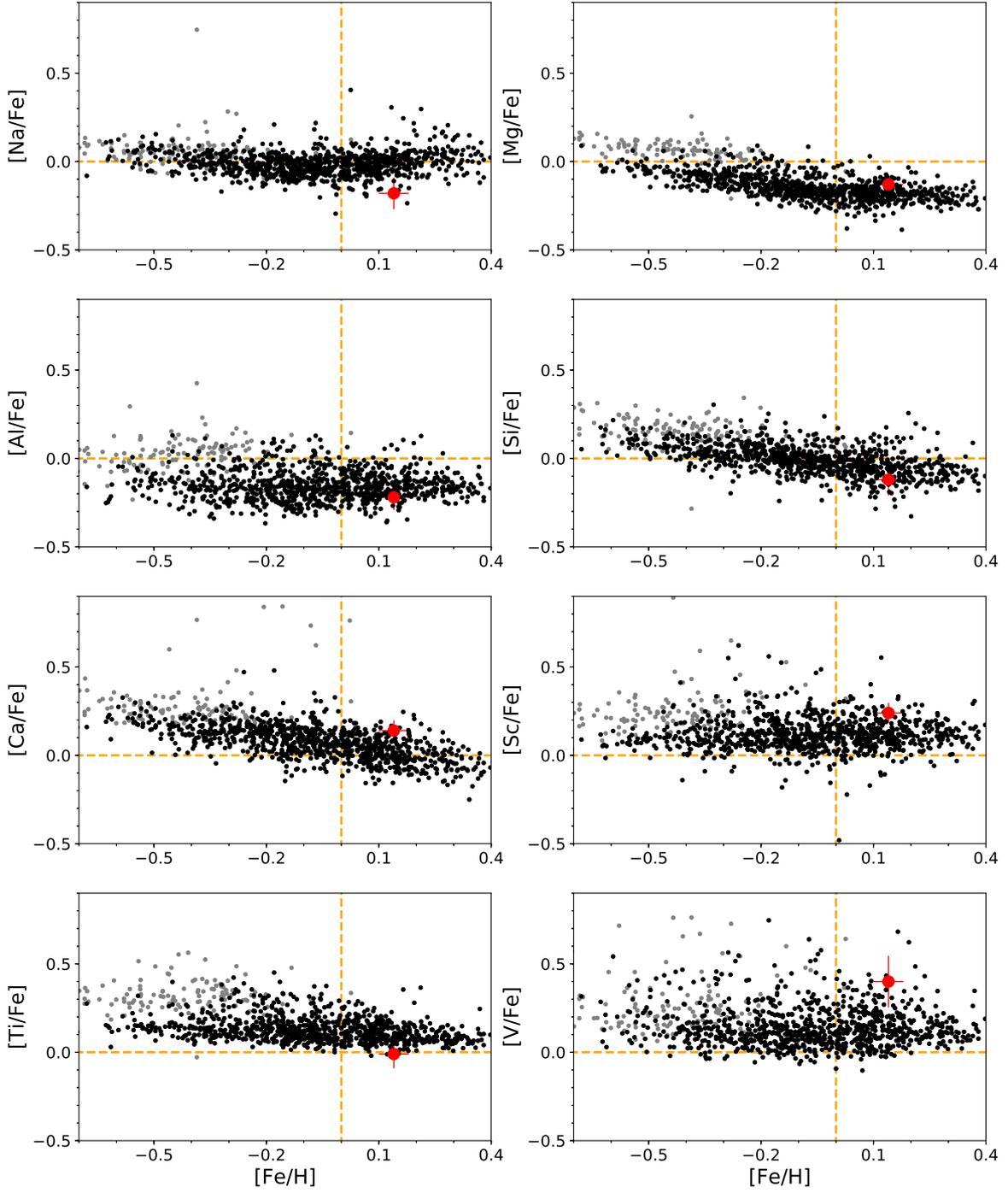}   
\vspace{-1.cm}
  \caption{Abundance ratios [X/Fe] versus [Fe/H]. The red circle shows the average value for ASCC\,123 and the error bars are the standard errors
  reported in Table~\ref{Tab:abund_cluster_sun}. The gray and black dots represent the galactic trends for the thick and thin disc, respectively, in the 
  solar neighborhood from the internal data release 4 of the \gaia-ESO Survey. 
  The dashed lines indicate the solar value. For all the elements the same scale is used for a proper comparison. 
  } 
  \label{trends1_ascc} 
\end{figure*} 

\begin{figure*} 
  \centering         
\vspace{1.cm}
  \includegraphics[width=18cm]{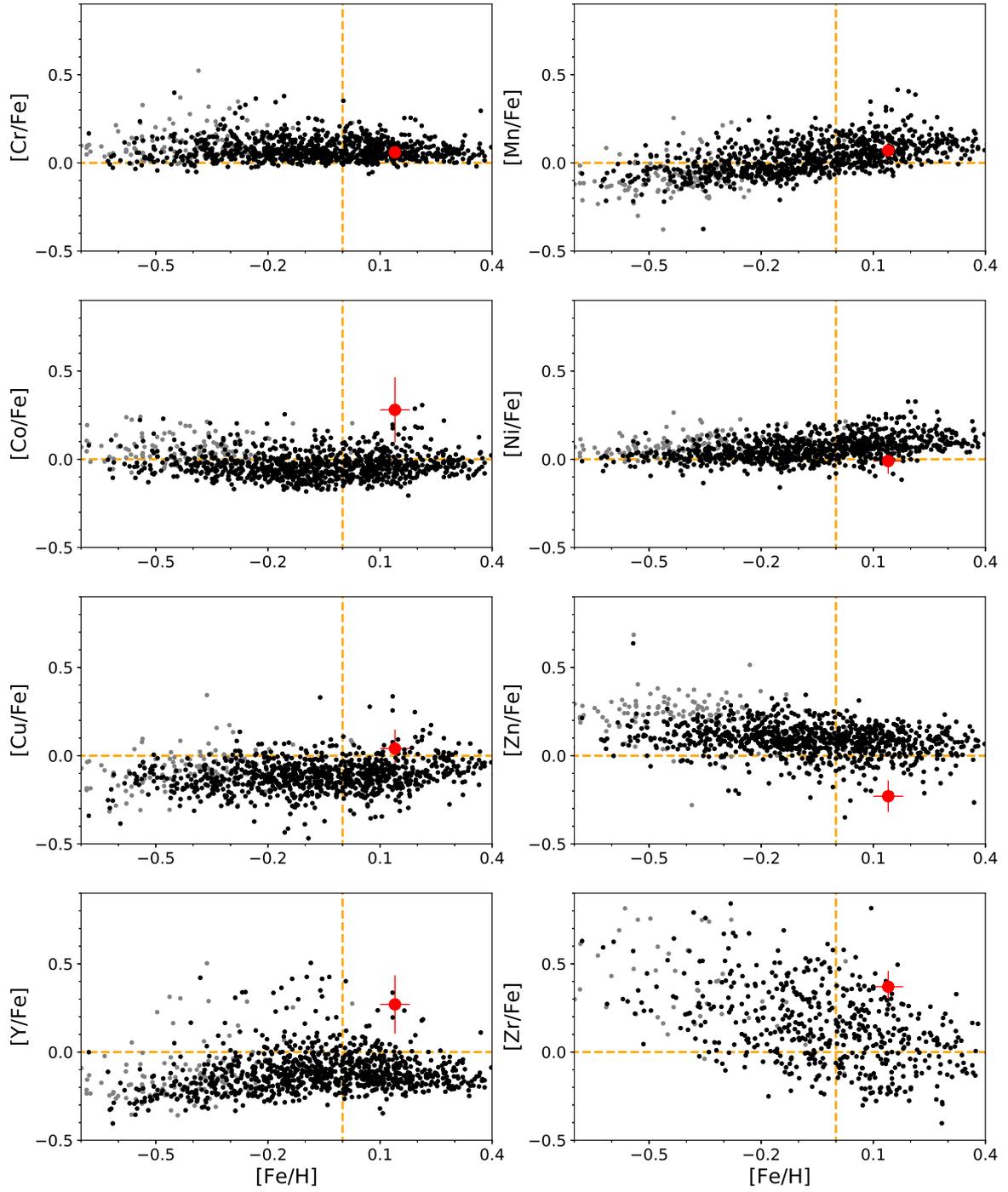}  
\vspace{-1.cm}
  \caption{Abundance ratios [X/Fe] versus [Fe/H].  Symbols are as in Fig.\,\ref{trends1_ascc}.} 
  \label{trends2_ascc}  
\end{figure*}

\end{document}